\documentclass[%
 reprint,
superscriptaddress,
 amsmath,amssymb,
 aps,prc,
]{revtex4-2} 
\usepackage{multirow}
\usepackage{graphicx}
\usepackage{dcolumn}
\usepackage{bm}
\usepackage{upgreek}
\usepackage[hidelinks]{hyperref}
\usepackage{CJKutf8}
\usepackage{longtable}
\begin{document}
\preprint{APS/PLACEHOLDER}
\title{Intruder structures in $^{32}$Si and $^{29}$Al}

\author{J.~Williams} 
 \email{ewilliams@triumf.ca}
\affiliation{TRIUMF, 4004 Wesbrook Mall, Vancouver, British Columbia, Canada V6T 2A3}

\author{G.~Hackman} 
\affiliation{TRIUMF, 4004 Wesbrook Mall, Vancouver, British Columbia, Canada V6T 2A3}
\affiliation{Department of Chemistry, Simon Fraser University, 8888 University Drive, Burnaby, British Columbia, Canada V5A 1S6}

\author{K.~Starosta} 
\affiliation{Department of Chemistry, Simon Fraser University, 8888 University Drive, Burnaby, British Columbia, Canada V5A 1S6}

\author{R.~S.~Lubna} 
\affiliation{Facility for Rare Isotope Beams, Michigan State University, 640 South Shaw Lane, East Lansing, MI 48824}

\author{Priyanka~Choudhary} 
\altaffiliation[Present address: ]{Department of Physics, KTH Royal Institute of Technology, Roslagstullsbacken 21 SE-106 91, Stockholm, Sweden}
\author{Subhrajit~Sahoo} 
\author{P.~C.~Srivastava} 
\affiliation{Department of Physics, Indian Institute of Technology Roorkee, Roorkee, India 247667}

\author{C.~Andreoiu}
\author{D.~Annen}
\affiliation{Department of Chemistry, Simon Fraser University, 8888 University Drive, Burnaby, British Columbia, Canada V5A 1S6}
\author{H.~Asch}
\affiliation{Department of Physics, Simon Fraser University, 8888 University Drive, Burnaby, British Columbia, Canada V5A 1S6}

\author{M.~D.~H.~K.~G.~Badanage}
\affiliation{Department of Chemistry, Simon Fraser University, 8888 University Drive, Burnaby, British Columbia, Canada V5A 1S6}

\author{G.~C.~Ball}
\affiliation{TRIUMF, 4004 Wesbrook Mall, Vancouver, British Columbia, Canada V6T 2A3}

\author{M.~Beuschlein}
\affiliation{Technische Universit{\"a}t Darmstadt, Department of Physics, Institute for Nuclear Physics, Schlossgartenstr. 9, 64289 Darmstadt, Germany}

\author{H.~Bidaman}
\author{V.~Bildstein}
\author{R.~J.~Coleman}
\affiliation{Department of Physics, University of Guelph, 50 Stone Rd E, Guelph, Ontario, Canada N1G 2W1}

\author{A.~B.~Garnsworthy}
\affiliation{TRIUMF, 4004 Wesbrook Mall, Vancouver, British Columbia, Canada V6T 2A3}

\author{B.~Greaves}
\affiliation{Department of Physics, University of Guelph, 50 Stone Rd E, Guelph, Ontario, Canada N1G 2W1}

\author{G.~Leckenby}
\altaffiliation[Present address: ]{LP2i Bordeaux, CNRS, Universit\'e de Bordeaux, F-33170 Gradignan, France}
\affiliation{TRIUMF, 4004 Wesbrook Mall, Vancouver, British Columbia, Canada V6T 2A3}
\affiliation{Department of Physics and Astronomy, University of British Columbia, Vancouver, British Columbia, Canada V6T 1Z1}

\author{V.~Karayonchev} 
\altaffiliation[Present address: ]{Argonne National Laboratory, 9700 S. Cass Avenue, Lemont, IL 60439}
\affiliation{TRIUMF, 4004 Wesbrook Mall, Vancouver, British Columbia, Canada V6T 2A3}

\author{M.~S.~Martin}
\altaffiliation[Present address: ]{Argonne National Laboratory, 9700 S. Cass Avenue, Lemont, IL 60439}
\affiliation{Department of Physics, Simon Fraser University, 8888 University Drive, Burnaby, British Columbia, Canada V5A 1S6}

\author{C.~Natzke}
\affiliation{TRIUMF, 4004 Wesbrook Mall, Vancouver, British Columbia, Canada V6T 2A3}

\author{C.~M.~Petrache}
\affiliation{Universit\'{e} Paris-Saclay, CNRS/IN2P3, IJCLab, 91405 Orsay, France}

\author{A.~Radich}
\affiliation{Department of Physics, University of Guelph, 50 Stone Rd E, Guelph, Ontario, Canada N1G 2W1}

\author{E.~Raleigh-Smith} 
\affiliation{TRIUMF, 4004 Wesbrook Mall, Vancouver, British Columbia, Canada V6T 2A3}

\author{D.~Rhodes}
\altaffiliation[Present address: ]{Lawrence Livermore National Laboratory, 7000 East Avenue, Livermore, CA 94550}
\affiliation{TRIUMF, 4004 Wesbrook Mall, Vancouver, British Columbia, Canada V6T 2A3}

\author{R.~Russell}
\affiliation{University of Surrey, Guildford, United Kingdom GU2 7XH}

\author{M.~Satrazani}
\altaffiliation[Present address: ]{KU Leuven, Instituut voor Kern- en Stralingsfysica, Leuven, Belgium}
\affiliation{University of Liverpool, Liverpool, United Kingdom L69 3BX}

\author{P.~Spagnoletti}
\altaffiliation[Present address: ]{Department of Physics, University of Liverpool, Liverpool, United Kingdom L69 7ZE}
\affiliation{Department of Chemistry, Simon Fraser University, 8888 University Drive, Burnaby, British Columbia, Canada V5A 1S6}

\author{C.~E.~Svensson} 
\affiliation{TRIUMF, 4004 Wesbrook Mall, Vancouver, British Columbia, Canada V6T 2A3}
\affiliation{Department of Physics, University of Guelph, 50 Stone Rd E, Guelph, Ontario, Canada N1G 2W1}

\author{D.~Tam}
\affiliation{Department of Physics, Simon Fraser University, 8888 University Drive, Burnaby, British Columbia, Canada V5A 1S6}
\author{F.~Wu (\begin{CJK*}{UTF8}{gbsn}吴桐安\end{CJK*})}
\affiliation{Department of Chemistry, Simon Fraser University, 8888 University Drive, Burnaby, British Columbia, Canada V5A 1S6}

\author{D.~Yates}
\affiliation{TRIUMF, 4004 Wesbrook Mall, Vancouver, British Columbia, Canada V6T 2A3}
\affiliation{Department of Physics and Astronomy, University of British Columbia, Vancouver, British Columbia, Canada V6T 1Z1}

\author{Z.~Yu}
\affiliation{Department of Chemistry, Simon Fraser University, 8888 University Drive, Burnaby, British Columbia, Canada V5A 1S6}

\date{\today}
\begin{abstract}
We have studied $^{32}$Si and $^{29}$Al using $^{12}$C($^{22}$Ne,2p) and $^{12}$C($^{22}$Ne,$\alpha$p) fusion-evaporation reactions.  In both cases, we observed significant population of high-spin structures distinct from the ground-state yrast bands.  In $^{32}$Si, most of the high-energy states feed into a $J^{\pi} = 5^-$ nanosecond isomer.  In $^{29}$Al, we identified a rotor-like negative-parity band with a $J^{\pi} = 7/2^-$ band-head. Doppler shift lifetime measurements were performed for all observed states. These results were compared to shell model calculations and interpreted in terms of proton and neutron cross-shell excitation.
\end{abstract}
\maketitle

\section{\label{sec:intro}Introduction}

\par Nuclides near the $N=20$ `island of inversion' contain important information on nuclear structure.  In the neutron-rich Na and Mg isotopes, residual nucleon-nucleon interactions lead to the disappearance of the $N=20$ shell gap and significant occupation of neutron $fp$-shell orbitals in ground-state configurations \cite{otsuka_tensorforce,ioi_theory,forces_utsuno}. In the nearby $sd$ shell region outside of the `island of inversion', the evolution of the $N=20$ shell closure is indicated by intermediate-energy negative-parity states which arise mainly due to single-neutron excitation to the higher-lying $fp$ orbitals.  Identification of these negative-parity states is important for mapping $N=20$ shell evolution.  For many nuclides in this region, high-spin yrast states are expected to have negative parity, since high-spin configurations will typically involve the $0f_{7/2}$ intruder orbital.  These high-spin intruder states should be strongly populated via fusion-evaporation reactions, and would be then expected to decay to lower-lying negative-parity states and eventually to positive-parity states in the ground-state band, allowing access to both positive and negative-parity states in the intermediate energy regime.

\par We have investigated high-spin states in the neutron-rich $sd$-shell nuclides $^{32}$Si and $^{29}$Al using fusion-evaporation reactions. In $^{32}$Si, which has not been previously studied using fusion-evaporation, we found several high-lying intruder states which feed a $J^{\pi} = 5^-$ isomer previously characterized in Ref.~\cite{williams_2023}, as well as promptly decaying high spin positive-parity states.  In $^{29}$Al, further examination of an excited band previously identified in Refs.~\cite{dungan_2016, sultana_2018} suggests that it has negative parity, and we have also identified several new high-energy states.  Our results show that in both $^{32}$Si and $^{29}$Al, the high-spin yrast states are negative parity and have significant $0f_{7/2}$ population. These states decay to the lower spin yrast states by hindered \emph{E1} or \emph{E3} transitions, in agreement with shell model predictions.

\section{Experimental details}
\label{sec:expt}

\par The nuclei of interest were populated using $^{12}$C($^{22}$Ne,2p) and $^{12}$C($^{22}$Ne,$\alpha$p) reactions with a $^{22}$Ne beam energy of 2.56A~MeV. Estimated cross sections for various fusion-evaporation exit channels observed in our data are listed in Table \ref{tab:xsec}.  $\gamma$ rays were detected using the TRIUMF-ISAC Gamma Ray Escape Suppressed Spectrometer (TIGRESS) \cite{tigress} instrumented with 14 Compton suppressed segmented high-purity germanium clover detectors --- 4 each at 45$^{\circ}$ and 135$^{\circ}$, and 6 at 90$^{\circ}$. The high peak-to-total configuration of TIGRESS was used, with the clover detectors positioned 14.5 cm from the target position. Clover addback was used to improve the photopeak-to-total efficiency of the array at high energies.  Relative intensities of $\gamma$ rays were corrected for summing using the 180$^{\circ}$ coincidence method of Ref.~\cite{griffin2019}. Energy and absolute efficiency calibrations of TIGRESS were performed using $^{56,60}$Co, $^{133}$Ba, and $^{152}$Eu sources, covering an energy range of 80 - 3451~keV (the fit function was extrapolated outside of this range). Charged particles were detected using a 128-channel spherical array of CsI(Tl) scintillators \cite{csiball}. Charged-particle identification was performed using offline pulse-shape analysis \cite{tip} to separate the exit channels populating $^{32}$Si and $^{29}$Al.  Fusion-evaporation events were further separated from the background using a series of timing gates for TIGRESS-TIGRESS hits ($\pm$60 ns), CsI-CsI hits ($\pm$200 ns), and TIGRESS-CsI hits ($\pm$240 ns). A thin self-supporting 500 $\upmu$g/cm$^2$ $^{nat.}$C foil produced by Micromatter \cite{micromatter} was used with a 23.6 mg/cm$^2$ lead catcher foil mounted approximately 1 mm downstream.  This catcher foil was used to stop recoils prior to the decay of any isomeric states, allowing for separation of prompt and isomeric transitions based on the Doppler shift.  An alternate target was used for lifetime measurements of short-lived states using the Doppler Shift Attenuation Method (DSAM), consisting of a 394 $\upmu$g/cm$^2$ layer of $^{nat.}$C mechanically rolled onto a 24 mg/cm$^2$ lead backing with an intermediate 200 $\upmu$g/cm$^2$ indium `sticking' layer to ensure adherence between the carbon and lead \cite{targets}.

\begin{table}
\caption{Approximate cross sections for fusion-evaporation exit channels observed in this work. The absolute efficiency of TIGRESS was determined from source data, while a value of 65\% was assumed for charged particles (based on Ref.~\cite{csiball}). Values assume 75\% population of the level decaying by the strongest observed transition with energy $E_{\gamma}$ listed below.}
\smallskip
\centering
\begin{ruledtabular}
\begin{tabular}{llr}
\rule{0pt}{2.5ex}Reaction ($E_{beam}=56.4$ MeV) & $E_{\gamma}$ (keV) & $\sigma$ (mb) \\ \hline
\rule{0pt}{2.5ex}$^{12}$C($^{22}$Ne,2p)$^{32}$Si  & 1942.06(6) & 0.61 \\
$^{12}$C($^{22}$Ne,$\alpha$p)$^{29}$Al            & 1754.3(4) & 3.24 \\
$^{12}$C($^{22}$Ne,$2 \alpha$)$^{26}$Mg           & 1808.6(5) & 12.77
\end{tabular}
\end{ruledtabular}
\label{tab:xsec}
\end{table}

\par Spins of populated states were determined from $\gamma$-ray angular distributions and directional correlations (DCO) \cite{dco}.  Angular distributions were fit to the function

\begin{equation}
W(\theta) = A_0(1 + a_2P_2(\text{cos }\theta) + a_4P_4(\text{cos }\theta)),
\label{eq:angdist}
\end{equation}

\noindent where $P_L(\text{cos }\theta)$ are Legendre polynomials, and the signs of $a_2$ and $a_4$ can be used to distinguish between dipole and quadrupole transitions ($a_2 > 0$, $a_4 < 0$ for quadrupole, $a_2 < 0$, $a_4 \approx 0$ for dipole, other values for mixed transitions). The data was grouped into 6 bins in $\theta$, based on the TIGRESS segment angles summed about $90^{\circ}$, in order to account for any differences in $\gamma$-ray attenuation at upstream and downstream angles due to the geometry of the target assembly. The relative efficiency of TIGRESS in each bin was determined using the calibration source data. For $\gamma-\gamma$ cascades where sufficient statistics were available, the DCO ratio $R_{DCO}$ was also measured:

\begin{equation}
R_{DCO} = \frac{I_{\gamma 2, \theta 1} (\text{Gated on }\gamma_1\text{ at }\theta_2)}{I_{\gamma 2, \theta 2} (\text{Gated on }\gamma_1\text{ at }\theta_1)}.
\end{equation}

The ratio $R_{DCO}$ has a value of $\sim$1.0 for stretched quadrupole transitions, $\sim$0.5 for stretched dipole, and other values for mixed transitions, when gating on a stretched quadrupole transition.  The two angles $\theta_1$ and $\theta_2$ were assigned to the 6 TIGRESS clovers at 90$^{\circ}$, and the 8 TIGRESS clovers at 45$^{\circ}$ or 135$^{\circ}$, respectively.  The validity of the angular distribution and DCO ratio analysis was tested using transitions of known multipolarity in the $^{26}$Mg channel.

\par Transitions with sufficient statistics were investigated using the polarization-direction correlation method outlined in Ref.~\cite{STAROSTA199916} to determine the parity of the parent or daughter state. For events where a $\gamma$-ray undergoes Compton scattering and is detected within two crystals of a TIGRESS clover, the scattering direction may be determined relative to the reaction plane defined by the beam axis and the initial direction of the $\gamma$-ray. The asymmetry $\Delta_{asym}$ in the scattering direction depends on the electric or magnetic character of the transition:

\begin{equation}
\Delta_{asym} = \frac{(a(E_{\gamma})N_{\perp}) - N_{\parallel}}{(a(E_{\gamma})N_{\perp}) + N_{\parallel}},
\end{equation}

\noindent where $N_{\perp}$ is the number of Compton-scattering events perpendicular (taken as $75^{\circ} \leq \theta \leq 105^{\circ}$ in this analysis) to the reaction plane, $N_{\parallel}$ is the number of Compton scattering events parallel to the reaction plane (taken as $\theta \leq 15^{\circ}$ or $\theta \geq 165^{\circ}$).  The correction factor $a(E_{\gamma})$ is based on the scattering direction asymmetry measured for unpolarized photons:

\begin{equation}
a(E_{\gamma}) = \frac{N_{\parallel}(\text{unpolarized})}{N_{\perp}(\text{unpolarized})},
\end{equation}

\noindent which was measured as a function of total energy $E_{\gamma}$ using $^{56}$Co source data.  A very small energy dependence was found for $a$, which varied from 0.94 to 0.92 in the energy range 1-4~MeV.

\par In cases where mixed \emph{M1/E2} multipolarity could be established and sufficient statistics were available, the \emph{M1/E2} mixing ratio $\delta$ was measured using the angular distribution data.  The angular distribution is affected by the mixing ratio, as well as the de-orientation parameter $\sigma$ which is affected by the reaction and de-excitation which occurs prior to the transition of interest, and the finite size of the detectors. The latter effect was estimated using simulations based on the GEANT4 framework  \cite{geant4_2}.  The de-orientation parameter $\sigma$ was either fit as a free parameter alongside $\delta$, or was measured based on the angular distributions of other transitions de-exciting the same level (eg.~stretched \emph{E2} transitions where $\delta = 0$ could be assumed, leaving $\sigma$ as the only free parameter).

\par Lifetimes of populated states were determined from a comparison of the DSAM target lineshape data to GEANT4-based simulations, as described in Ref.~\cite{22ne}.
For states with minimal or no observed feeding, the `effective' lifetime without feeding corrections (which is potentially longer than the true lifetime) was reported. For states with significant observed feeding, a feeding correction was applied by gating on the energy of a feeding transition, and comparing the coincident lineshape of the lower transition to simulations incorporating the measured effective lifetime of the feeding. The Compton scattering background underneath the lineshape was fit to a 2$^{\text{nd}}$ order polynomial function. In cases where background peaks were identified near the lineshape of interest, these were subtracted by placing an energy gate on a known coincident $\gamma$ ray, and (when necessary) gating on the Compton background near the coincident $\gamma$ ray and subtracting the resulting background spectrum from the lineshape. The systematic uncertainty due to the electronic stopping powers was evaluated by simulating lineshapes using both the default GEANT4 stopping powers based on ICRU Report 73 \cite{icru73} and stopping powers from SRIM-2013 \cite{srim}, with the average of best-fit lifetimes using these two methods taken as the adopted lifetime, and the difference taken as the uncertainty due to the stopping process. The average of these stopping models has been shown to agree well with measured stopping powers in this mass region and in the recoil energy range of interest \cite{williams25na}.

\par
\begin{figure}
\begin{center}
\includegraphics[width=1.0\columnwidth]{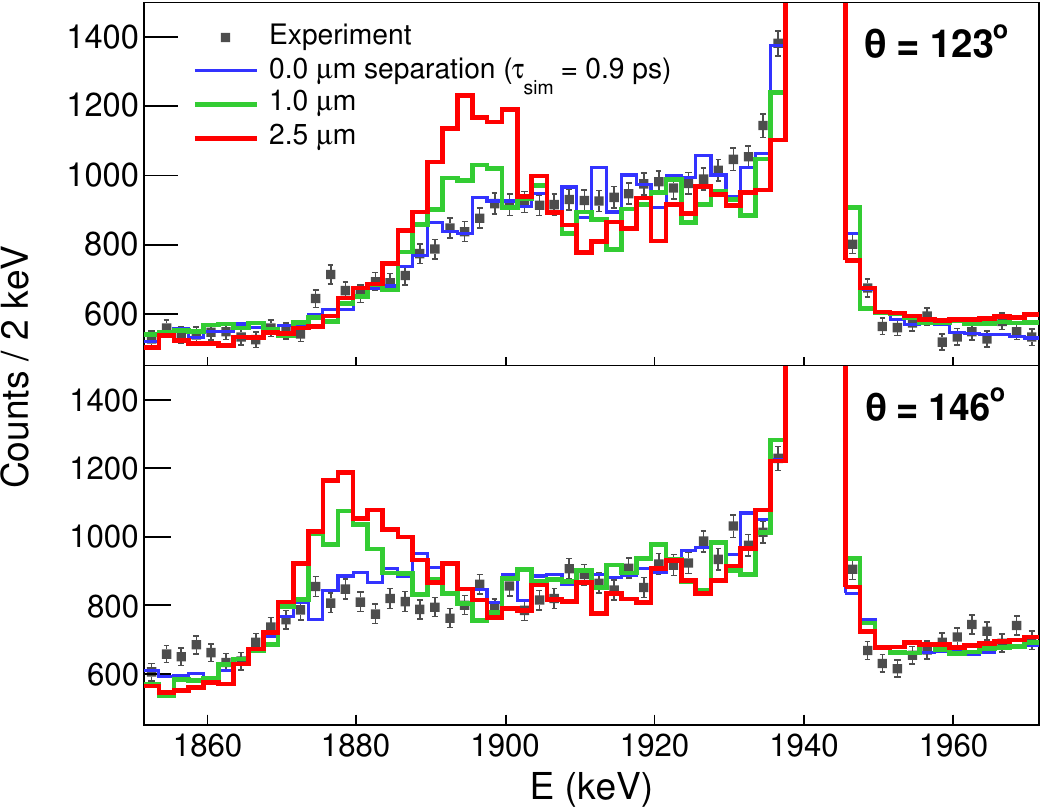}
\end{center}
\caption{Comparison of two-proton gated $\gamma$ singles DSAM target data in two TIGRESS segment rings to simulated lineshapes for various separation distances between the indium and lead target layers.}
\label{fig:target_gap_comp}
\end{figure}

 A potential concern for our lifetime measurements is delamination of the lead and indium layers in the DSAM target, which were originally coupled by mechanical rolling \cite{targets}. To investigate this effect, we compared our DSAM target data to simulated lineshapes which assumed various separation distances between the lead and indium layers. The $\gamma$ singles data for the $2^+_1 \rightarrow 0^+_1$ transition in $^{32}$Si was used for this comparison due to its high statistics and the presence of a Doppler shifted component in its lineshape which would be affected by target delamination. Since the stopped component contains feeding from a known isomer \cite{williams_2023}, this feeding was simulated separately and added to the lineshape. The amplitudes of the prompt-fed and isomer-fed components were allowed to vary freely in the fit, which removed the sensitivity to the level lifetime, but allowed for the best possible fit of the Doppler shifted portion of the lineshape that is affected by target delamination. The simulations used target layer separation values of 0.0 - 2.5 $\upmu$m and feeding-uncorrected lifetimes in the range $0.6 \leq \tau_{sim} \leq 1.0$ ps, and the optimal target layer separation for each lifetime was determined using the $\chi^2$ minimization method of Ref.~\cite{22ne}. The target layer separation was consistent with zero for all $\tau_{sim}$ values: an upper limit of 0.01 $\upmu$m (at 90\% confidence) was obtained from the simulations at $\tau_{sim} = 0.9$ ps, and smaller limits were obtained for all other $\tau_{sim}$ values. A comparison of simulated lineshapes to the DSAM target data is shown in Figure \ref{fig:target_gap_comp}. These simulations indicate that that when there is a significant gap between target layers, a secondary peak is expected in the Doppler shifted component of the lineshape, corresponding to the decay of the recoiling nucleus as it traverses the gap. This feature is not visible in our data, suggesting that there is no significant delamination of the DSAM target.

\section{$^{32}$Si Data}
\label{sec:analysis_32si}

\par
\begin{figure*}
\begin{center}
\includegraphics[width=1.8\columnwidth]{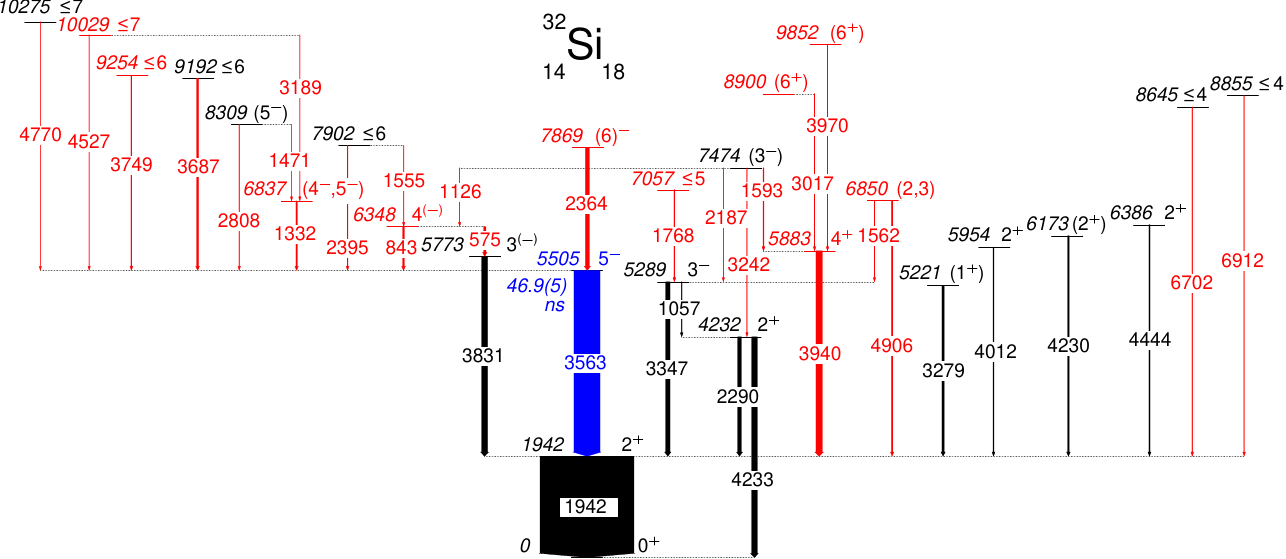}
\end{center}
\caption{Decay scheme of $^{32}$Si containing levels and transitions observed in this work.  Line widths indicate relative intensities of each transition.  Newly observed levels and transitions are in red, isomeric transitions in blue. Some spin values are reported as upper limits.}
\label{fig:32si_decayscheme}
\end{figure*}

\begin{figure}
\begin{center}
\includegraphics[width=1.0\columnwidth]{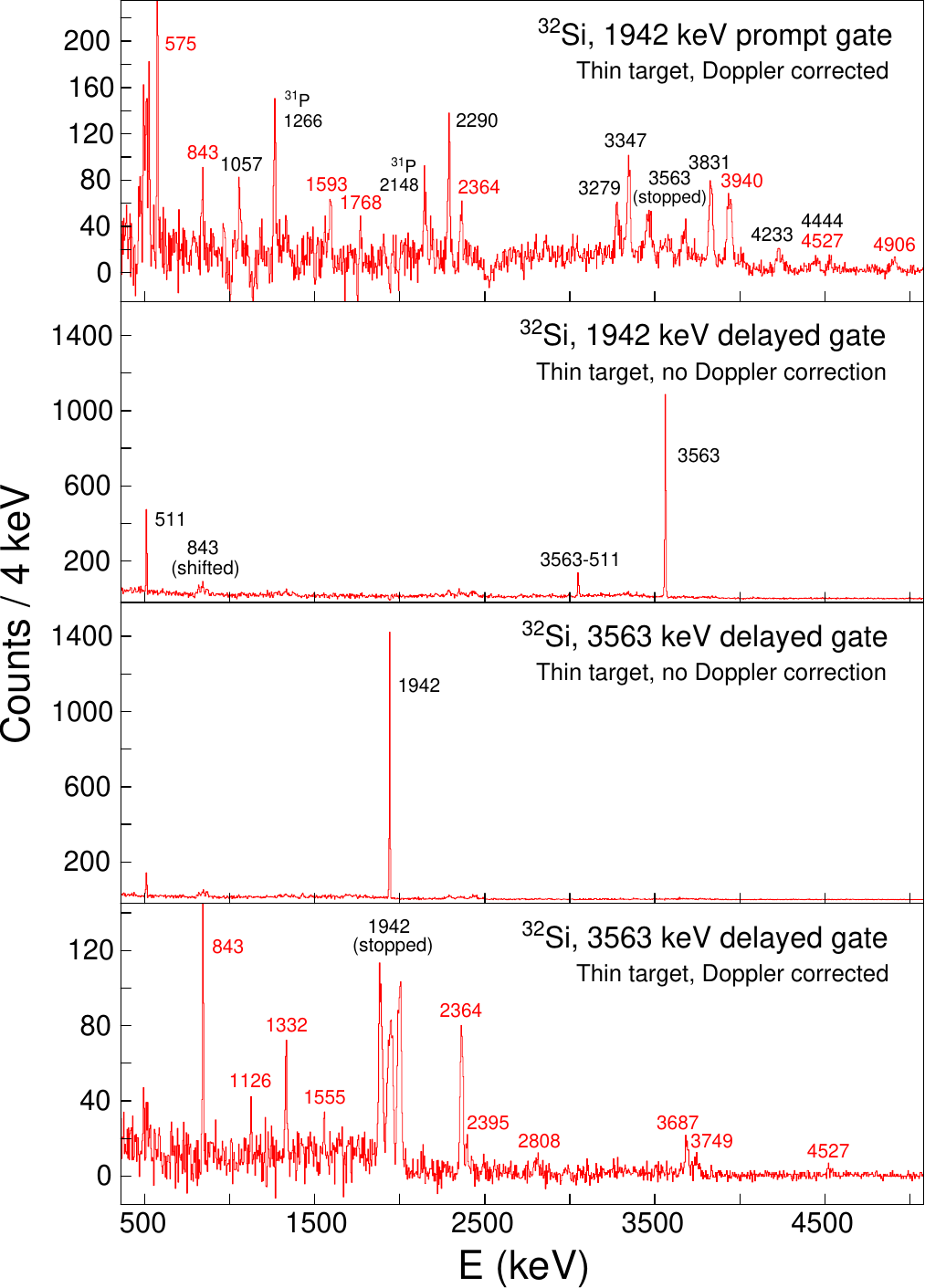}
\end{center}
\caption{Background-subtracted Doppler-corrected $\gamma$-ray spectra gated on the 1942.06(6)~keV and 3562.64(12)~keV transitions in $^{32}$Si.  Previously unobserved transitions are labelled in red.  A `prompt' gate indicates a gate on the Doppler shifted component of the specified transition, whereas a `delayed' gate indicates a gate on the stopped component (corresponding to decays of the recoiling nucleus after it was implanted in the downstream stopper foil).}
\label{fig:EDopp_spectra_32si}
\end{figure}

\begin{figure}
\begin{center}
\includegraphics[width=1\columnwidth]{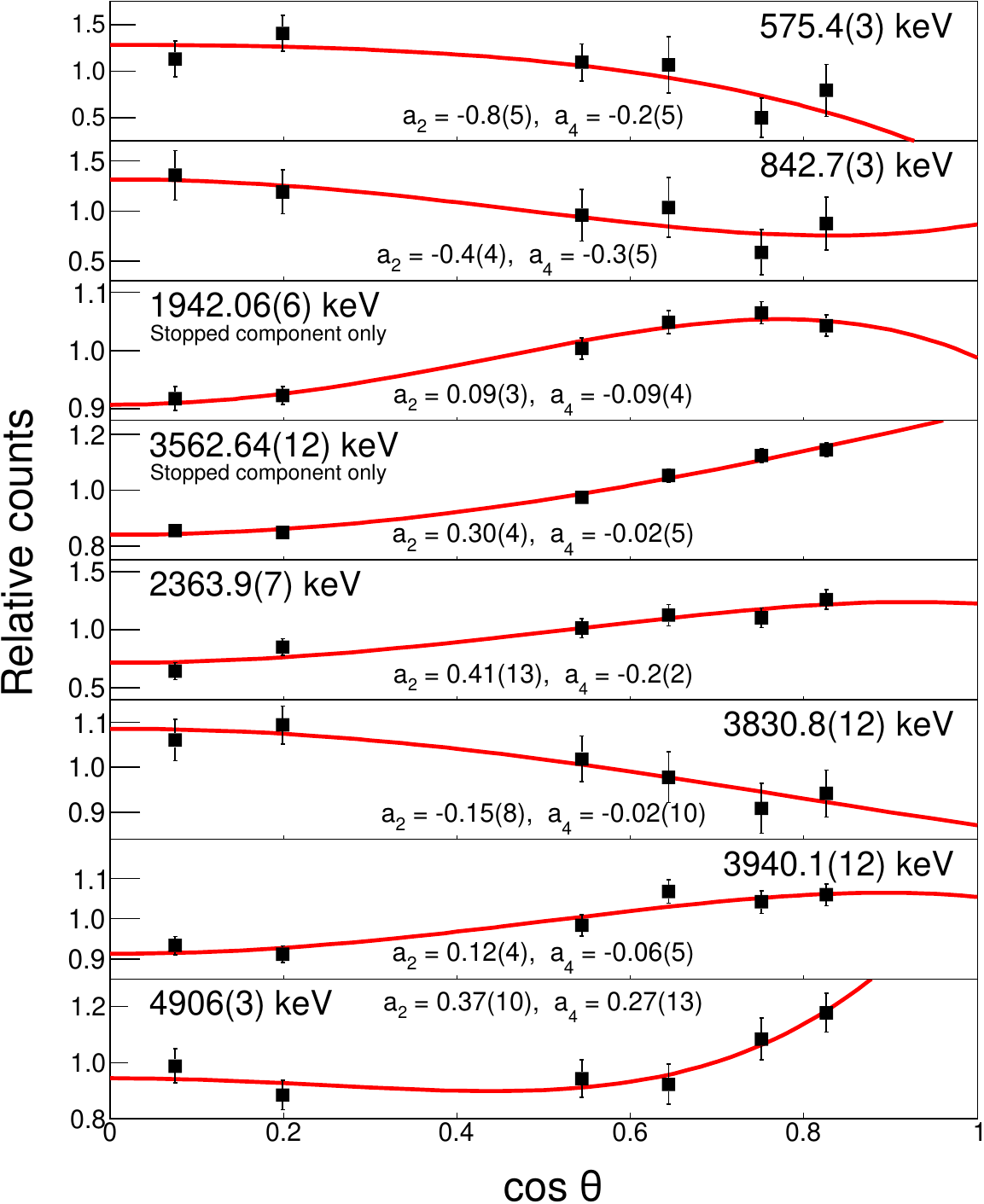}
\end{center}
\caption{Angular distributions of various transitions in $^{32}$Si, with fits to Eq. \ref{eq:angdist}.}
\label{fig:32Si_angdist}
\end{figure}

 A list of observed levels and transitions in $^{32}$Si is shown in Table \ref{tab:32si_levels}, and the corresponding decay scheme in Figure \ref{fig:32si_decayscheme}. Due to an improved fitting procedure, the $\gamma$-ray energies differ slightly from and supersede the values originally reported in Ref.~\cite{williams_2023}. Various Doppler-corrected and/or $\gamma$-gated TIGRESS spectra are shown in Figure \ref{fig:EDopp_spectra_32si}.  Angular distribution fit coefficients, $R_{DCO}$ values, and polarization asymmetry coefficients for select transitions are reported in Table \ref{tab:gamma_dist_table}, and several angular distribution fits are shown in Figure \ref{fig:32Si_angdist}. Detailed information on specific levels and their associated transitions follows:

\begin{table}
\caption{List of $^{32}$Si levels and $\gamma$ rays observed in this work.  Items in {\bf bold} are newly observed or measured ($I_{\gamma}$ values resulting in new branching ratios are also highlighted).  Lifetime limits are reported to 90\% confidence, all other quoted uncertainties are at 1$\sigma$ confidence.}
\smallskip
\centering
\begin{ruledtabular}
\begin{tabular}{llllr}
\rule{0pt}{2.5ex}$E_{level}$ (keV) & $E_{\gamma}$ (keV) & $I_{\gamma, rel}$ & $\tau_{mean}$ (fs) & $J^{\pi}$ \\ \hline
\rule{0pt}{2.5ex}1942.12(6)       & 1942.06(6)       & 100.0(5) & 780(120) & $2^+$\\
                 4232.4(7)        & 4232.6(13)        & 5.9(2) & 300(100)$^{\dag}$ & $2^+$ \\
                                  & 2290.0(8)        & 4.1(5) & & \\
                 5221.0(14)       & 3278.7(11)       & 2.3(3) & $< 56$ & $(1^+)$ \\ 
                 5288.8(9)        & 3346.5(11)       & 4.6(5) & 260(90) & $3^-$ \\
                                  & 1056.5(14)        & 0.8(2) & & \\
                 5504.88(13)      & 3562.64(12)      & 27.8(5) & $46.9(5) \times 10^6$ & $\mathbf{5^-}$ \\
                 5773.1(12)       & 3830.8(12)       & 6.2(2) & {\bf 40(30)} & $\mathbf{3^{(-)}}$ \\
                 {\bf 5882.5(12)} & {\bf 3940.1(12)} & 8.0(2) & {\bf 18(8)} & $\mathbf{4^+}$ \\
                 5954(3)          & 4012(3)          & 0.9(2) & $< 21$ & $2^+$ \\
                 6173(2)          & 4230(2)          & 1.7(3) & $< 43$ & $(2^+)$ \\ 
                 {\bf 6347.7(3)}  & {\bf 575.4(3)}   & {\bf 2.4(3)} & {\bf 980(140)} & $\mathbf{4^{(-)}}$ \\
                                  & {\bf 842.7(3)}   & {\bf 1.7(2)} & &  \\
                 6386(3)          & 4444(3)          & 1.1(2) & $< 126$ & $2^+$ \\
                 {\bf 6837.0(5)}  & {\bf 1332.0(5)}  & 1.3(3) & {\bf 50(40)}$^{\dag}$ & $\mathbf{(4^-,5^-)}$ \\
                 {\bf 6850(2)}    & {\bf 1562(2)}    & {\bf 0.38(12)} & $\mathbf{< 120}$ & $\mathbf{(2,3)}$ \\ 
                                  & {\bf 4906(3)}    & {\bf 1.4(2)} &  & \\ 
                 {\bf 7056.5(12)} & {\bf 1767.7(10)} & 0.9(2)     & $\mathbf{< 47}$  & $\mathbf{\leq 5}$ \\ 
                 7474.2(9)        & {\bf 1126.2(10)}  & {\bf 0.35(9)} & {\bf 80(30)}$^{\dag}$ & $\mathbf{(3^{-})}$ \\
                                  & {\bf 1592.5(13)} & {\bf 1.4(3)}   & & \\
                                  & {\bf 2187(2)}    & {\bf 0.32(10)} & & \\
                                  & {\bf 3242(2)}    & {\bf 0.5(2)} & & \\
                 {\bf 7868.9(7)}  & {\bf 2363.9(7)}  & 4.5(8) & {\bf 43(9)}$^{\dag}$ & $\mathbf{(6)^-}$ \\
                 7901.8(10)     & {\bf 1554.6(11)} & {\bf 0.53(13)} & {\bf 160(80)}$^{\dag}$ &  $\mathbf{\leq 6}$ \\
                                & {\bf 2395(2)}    & {\bf 0.8(2)} & & \\
                 8309.0(14)     & {\bf 1471(2)} & {\bf 0.33(9)} & {\bf 50(20)}$^{\dag}$ & ($5^-$) \\
                                & {\bf 2808(3)}    & {\bf 0.7(2)} & & \\
                 8645(5)        & {\bf 6702(5)}    & 0.8(2) & $\mathbf{< 320}$ & $\mathbf{\leq 4}$ \\
                 8855(7)         & {\bf 6912(7)}   & 0.6(2) & $\mathbf{< 103}$ & $\mathbf{\leq 4}$ \\
                 {\bf 8900(3)}    & {\bf 3017(3)}    & 0.7(3) & {\bf 110(40)}$^{\dag}$ & $\mathbf{(6^+)}$ \\
                 9192.2(14)          & {\bf 3687.0(14)}    & 2.0(4) & $\mathbf{< 39}$ & $\mathbf{\leq 6}$ \\
                 {\bf 9254(2)}    & {\bf 3749(2)}    & 1.3(3) & $\mathbf{< 12}$ & $\mathbf{\leq 6}$ \\
                 {\bf 9852(2)}    & {\bf 3970(2)}    & 1.24(13) & $\mathbf{< 110}$ & $\mathbf{(6^+)}$ \\
                 {\bf 10029(2)}   & {\bf 3189(3)}    & {\bf 0.33(10)} & $\mathbf{< 25}$ & $\mathbf{\leq 7}$ \\
                                  & {\bf 4527(3)}    & {\bf 0.38(10)} &  & \\
                 10275(4)         & {\bf 4770(4)}    & 0.19(6) & $\mathbf{< 430}$ & $\mathbf{\leq 7}$
\end{tabular}
\end{ruledtabular}
{\footnotesize \rule{0pt}{2.5ex}$^{\dag}$Effective lifetime without any feeding correction.}
\label{tab:32si_levels}
\end{table}

\begin{table*}
\caption{List of $a_2$, $a_4$, $R_{DCO}$, and $\Delta_{asym}$ values measured for transitions in $^{32}$Si and $^{29}$Al, and inferred multipolarities.}
\smallskip
\centering
\begin{ruledtabular}
\begin{tabular}{lllllr|r}
\rule{0pt}{2.5ex}$E_{\gamma}$ (keV) & $a_2$ & $a_4$ & DCO gate $E_{\gamma}$ (keV) & $R_{DCO}$ & $\Delta_{asym}$ & Assignment \\ \hline
\rule{0pt}{2.5ex}{\bf $^{32}$Si} \\
575.4(3)    & -0.8(5)  & -0.2(5)  & 1942.06(6)  & 0.9(2)   & $\S$ & D+Q \\
842.7(3)    & -0.4(4)  & -0.3(5)  & 1942.06(6), 3562.64(12) & 0.6(2), 0.8(2) & $\S$ & D+Q \\ 
1126.2(10)  & -0.1(8)  & -0.3(10) & 3562.64(12) & 0.4(6)   & $\S$ & (D+Q) \\
1332.0(5)   & -0.4(5)  & -0.2(6)  & 3562.64(12) & 0.5(4)   & $\S$ & (D+Q) \\
1942.06(6)  & 0.09(3)  & -0.09(4) & $\ddag$ & $\ddag$ & -0.002(15) & Q \\
2290.0(8)   & -1.1(5) & -0.3(5)   & 1942.06(6)  & 0.8(2)   & $\S$ & D+Q \\
2363.9(7)   & 0.41(13) & -0.2(2) & 1942.06(6), 3562.64(12)  & 1.11(9), 1.11(11)  & $-0.14(8)$ & (M1+)E2 \\
2395(2)     & $\S$ & $\S$ & 3562.64(12) & 0.8(3)   & $\S$ & (D+Q) \\
2808(3)     & $\S$ & $\S$ & 3562.64(12) & 0.7(3)   & $\S$ & (D+Q) \\
3017(3)     & $\S$ & $\S$ & 3940.1(12)  & 1.7(6)   & $\S$ & (Q) \\
3278.7(11)  & 0.32(12)  & 0.0(2)    & 1942.06(6)  & 0.8(3)   & $+0.12(16)$  & (D+Q) \\
3346.5(11)  & -0.43(11) & -0.04(14) & 1942.06(6)  & 0.4(2)   & $+0.05(8)$   & D \\
3562.64(12) & 0.30(4) & -0.02(5) & 1942.06(6)  & 1.06(4)  & $+0.032(13)$ & E3$^{*}$ \\
3687.0(14)  & 0.1(4)  & 0.1(5)   & 3562.64(12) & 0.7(2)   & $\S$ & D(+Q) \\
3749(2)     & 0.7(6)  & -0.2(7)  & 3562.64(12) & 0.6(2)   & $\S$ & D(+Q) \\
3830.8(12)  & -0.15(8) & -0.02(10) & 1942.06(6) & 0.48(14) & $-0.01(3)$ & D \\
3940.1(12)  &  0.12(4) & -0.06(5) & 1942.06(6) & 1.00(9)  & $+0.07(3)$ & E2 \\
3970(2)     & 0.18(13) &  0.0(2)  & 1942.06(6), 3940.1(12) & 1.3(4), 1.7(6)   & $\S$ & (Q) \\
4232.6(13)  & -0.12(6) & 0.18(7)  & 1942.06(6)  & 0.5(3)   & $\S$ & D(+Q) \\
4906(3)     & 0.37(10) & 0.27(13) & 1942.06(6)  & 0.8(3)   & $\S$ & D+Q \\ \hline
\rule{0pt}{2.5ex}{\bf $^{29}$Al} \\
647.0(7)   & -0.21(3) & -0.07(3) & 3576.6(13) & 0.6(4) & $-0.098(15)$, $-0.13(5)^{\dag}$  & M1/E2 \\
704.0(7)   & -0.39(14) & -0.1(2) & 3576.6(13) & 0.8(3) & $-0.10(3)$, $-0.13(9)^{\dag}$ & M1/E2 \\
1212.3(5)  & -0.4(2)  & -0.2(3)  & 3576.6(13) & 0.9(2) & $-0.07(3)$, $-0.03(11)^{\dag}$ & M1/E2 \\
1292.3(5)  & -0.13(4) & -0.04(5) & 3576.6(13) & 0.9(2) & $-0.14(5)^{\dag}$ & M1/E2 \\
1412.7(5)  & -0.30(15) & -0.1(2) & 3576.6(13) & 0.9(4) & $\S$ & D(+Q) \\
1699.0(5)  & -0.44(12)  & -0.21(15)  & $\ddag$ & $\ddag$ & $\S$ & D(+Q) \\
1754.3(4)  & -0.067(8)  & -0.013(10) & 3687.9(12) & 0.6(2)   & $-0.039(3)$ & M1(+E2) \\
1823.4(4)  & -0.096(8)  & -0.022(10) & 3687.9(12) & 0.67(14) & $-0.028(3)$, $-0.056(18)^{\dag}$ & M1(+E2) \\
2277.4(5)  & -0.26(3)& -0.03(3) & 3576.6(13) & 0.9(2) & $-0.040(12)$, $-0.06(5)^{\dag}$ & M1(+E2) \\
2334.0(5)  & -0.38(4)& -0.10(4) & 3576.6(13) & 0.8(2) & $+0.015(13)$, $+0.05(4)^{\dag}$ & (E1) \\
2825.4(7)  & 0.09(11)& 0.07(14)& 3576.6(13) & 0.4(5) & $\S$ & (D+Q) \\
2952.6(13) & 0.0(2)  & 0.0(3)  & 3576.6(13) & 0.7(3) & $\S$ & N/A \\
2991.4(8)  & -0.25(13)& 0.2(2) & 3576.6(13) & 0.9(2) & $+0.03(3)$, $+0.07(7)^{\dag}$ & (D+Q) \\
3186.4(10) & -0.2(3) & -0.2(3) & 3576.6(13) & 1.1(6) & $+0.18(12)^{\dag}$ & (D) \\
3510.0(10) & -0.38(4)& -0.07(5)& $\ddag$ & $\ddag$ & $0.00(2)$, $+0.09(6)^{\dag}$ & (E1) \\
3576.6(13) & 0.18(4) & -0.07(5)& $\ddag$ & $\ddag$ & $+0.047(19)$ & E2 \\
3687.9(12) & 0.4(2)  & 0.0(2)  & 3576.6(13) & 1.9(6) & $\S$ & Q \\
3936.8(14) & -0.68(5)& 0.14(6) & $\ddag$ & $\ddag$ & $-0.035(14)$ & M1/E2 \\
3976.1(15) & 0.4(2)  & -0.3(3) & $\ddag$ & $\ddag$ & $\S$ & (Q) \\
4097(2)    & 0.5(2)  & 0.1(3)  & $\ddag$ & $\ddag$ & $\S$ & (Q) \\
4110.1(15) & -0.1(2) & 0.3(3)  & $\ddag$ & $\ddag$ & $\S$ & (D+Q) \\
4403.7(14) & 0.28(9) & 0.14(12)& $\ddag$ & $\ddag$ &$\S$ & (D+Q) \\
4773(2)    & 0.3(2)  & 0.2(2)  & $\ddag$ & $\ddag$ & $\S$ & (Q) \\
4822(2)    & 0.2(3)  & 0.8(4)  & $\ddag$ & $\ddag$ & $\S$ & (D+Q) \\
4941.6(17) & 0.18(5) & 0.01(7) & $\ddag$ & $\ddag$ & 0.00(2) & (Q) \\
5011(2)    & 0.3(2)  & 0.1(3)  & $\ddag$ & $\ddag$ & $\S$ & (Q) \\
5729(2)    & 0.02(15) & 0.1(2) & $\ddag$ & $\ddag$ & $\S$ & (D+Q) \\
\end{tabular}
\end{ruledtabular}
{\footnotesize \rule{0pt}{2.5ex}$\S$: Insufficient statistics. $\ddag$: No suitable DCO gate. $^*$From decay scheme and lifetime of parent state as discussed in Ref.~\cite{williams_2023}.
$^{\dag}$After gating on 1754.3(4) keV transition.}
\label{tab:gamma_dist_table}
\end{table*}

\begin{table*}
\caption{Mixing ratio $\delta$ and orientation parameter $\sigma/J_i$ values measured for transitions in $^{32}$Si and $^{29}$Al.  The mixing ratio convention of Krane and Steffen \cite{krane_determination_1970} is used.  Where applicable, $B(M1; J^{\pi}_i \rightarrow J^{\pi}_f)$ and $B(E2; J^{\pi}_i \rightarrow J^{\pi}_f)$ values are listed based on the lifetime measurements of Tables \ref{tab:32si_levels} and \ref{tab:29al_levels}. Transition strength uncertainties were calculated using the Monte Carlo method of the Java-RULER program \cite{javaruler}. All limits are reported to 90\% confidence.}
\smallskip
\centering
\begin{ruledtabular}
\begin{tabular}{lllllr|rr}
\rule{0pt}{2.5ex}$E_{\gamma}$ & $J^{\pi}_i $ & $J^{\pi}_f $ & Assumed & $\sigma/J_i$ & $\delta$ & $B(M1; J^{\pi}_i$ & $B(E2; J^{\pi}_i$ \\
(keV) &  & & $\lambda L$ &  &  & $\rightarrow J^{\pi}_f)$ (W.u.) & $\rightarrow J^{\pi}_f)$ (W.u.)\\ \hline
\rule{0pt}{2.5ex}{\bf $^{32}$Si} \\
3940.1(12) & $4^+$ & $2^+$ & E2 & 0.85(11) & 0$^{\dag}$ & N/A & $8^{+7}_{-3}$ \\
3562.64(12) (thin target) & $5^-$ & $2^+$ & E3 & 0.58(3) & 0$^{\dag}$ & N/A & N/A \\ 
3562.64(12) (DSAM target) & $5^-$ & $2^+$ & E3 & 0.41(2) & 0$^{\dag}$ & N/A & N/A \\ 
575.4(3) & $4^{(-)}$ & $3^{(-)}$ & M1/E2 & $\leq 0.4^*$ & $< 0.00$ & $< 0.15$ & Unbounded \\
842.7(3) & $4^{(-)}$ & $5^-$ & M1/E2 & $\leq 0.4^*$ & $> +0.13$ & $< 0.36$ & $> 3.6$ \\
2363.9(7) & $(6)^-$ & $5^-$ & M1/E2 & $\leq 0.4^*$ & +1.0(6) & $2.8^{+2.1}_{-1.3} \times 10^{-2}$ & $21^{+11}_{-15}$ \\ \hline
\rule{0pt}{2.5ex}{\bf $^{29}$Al} \\
1754.3(4) & $7/2^+$ & $5/2^+$ & M1/E2 & 0.560(14) & +0.139$^{+0.010}_{-0.011}$ & 0.34$^{+0.05}_{-0.04}$ & 10$^{+3}_{-2}$ \\
1823.4(4) & $9/2^+$ & $7/2^+$ & M1/E2 & 0.55(2) & +0.071$^{+0.010}_{-0.011}$ & 0.4$^{+0.5}_{-0.2}$ & 3$^{+4}_{-2}$\\
3576.6(13) & $9/2^+$ & $5/2^+$ & E2 & 0.56(4) & 0$^{\dag}$ & N/A & 2.8$^{+3.3}_{-1.3}$ \\
2334.0(5) & $11/2^-$ & $9/2^+$ & E1 & 0.31(2) & 0$^{\dag}$ & N/A & N/A \\
647.0(7)  & $11/2^-$ & $9/2^-$ & M1/E2 & 0.36$^{+0.04}_{-0.05}$ & $+0.04(3)$ & 0.54$^{+0.09}_{-0.07}$ & 10$^{+22}_{-9}$ \\
1292.3(5) & $13/2^-$ & $11/2^-$ & M1/E2 & $\leq 0.35^*$ & $+0.095(15)$ & 1.1$^{+0.5}_{-0.3}$ & 30$^{+18}_{-10}$ \\
1699.0(5) & $15/2^-$ & $13/2^-$ & M1/E2 & $\leq 0.30^*$ & $-0.02(4)$ & $> 0.91$ & Unbounded \\
\end{tabular}
\end{ruledtabular}
{\footnotesize \rule{0pt}{2.5ex}$^{\dag}$Assumed $\delta = 0$, only $\sigma/J_i$ was fit.  $^*$Assumed limit based on $\sigma/J_i$ measured for other transitions.}
\label{tab:mixing_ratio_table}
\end{table*}

\begin{figure*}
\begin{center}
\includegraphics[width=2.05\columnwidth]{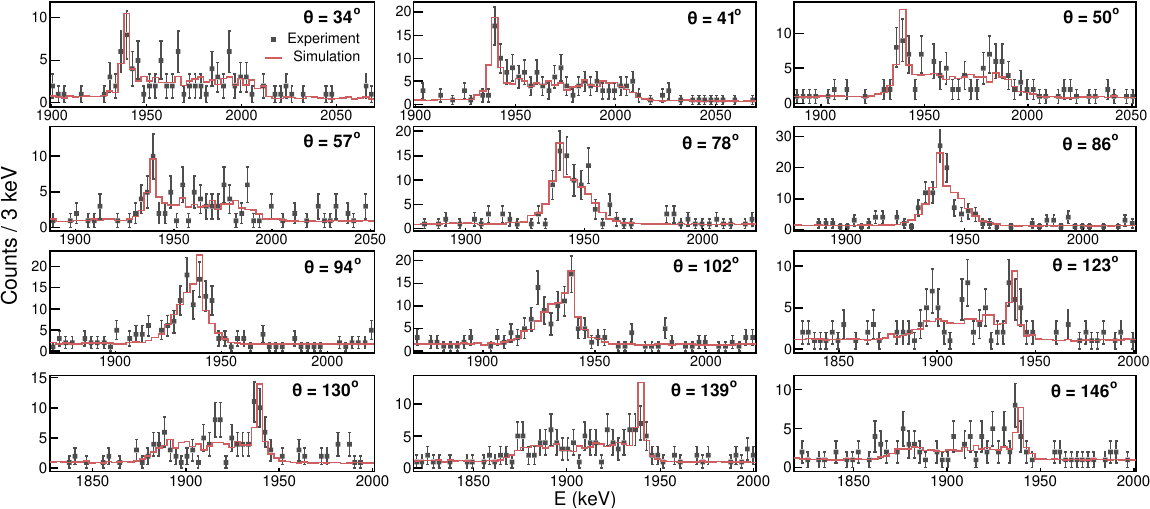}
\end{center}
\caption{Comparison of the 1942.06(6) keV DSAM lineshape (gated on the feeding 3940.1(12) keV transition) to GEANT4-based simulations assuming $\tau_{mean} = 800$ fs for the $2^+_1$ level of $^{32}$Si.}
\label{fig:lineshape_1942}
\end{figure*}

\par {\bf 1942 keV level:} Our angular distribution data agrees with a stretched quadrupole assignment for the 1942.06(6) keV ground-state transition. For the lifetime measurement, a gate from above on the Doppler shifted 3940.1(12)~keV transition was applied to constrain the effect of feeding on the DSAM lineshape.  This feeding was then incorporated into the lineshape simulation.  A comparison of the $\gamma$-gated data to the lineshape simulation is shown in Figure \ref{fig:lineshape_1942}. The best fit lifetime was 780(120) fs, resulting in a B($E2$; $2^+_1 \rightarrow 0^+_1$) value of 6.3$^{+1.1}_{-0.8}$ W.u.

\par {\bf 5505 keV isomer:} The decay scheme of $^{32}$Si is complicated by the presence of a nanosecond isomer, originally reported in Refs. \cite{32si_fornal} and \cite{32si_asai} with differing decay schemes.  We previously investigated this isomer in detail \cite{williams_2023}, assigning it $J^{\pi} = 5^-$, with $\tau_{mean} = 46.9(5)$ ns.  The isomer primarily decays by an \emph{E3-E2} $\gamma$ cascade with energies of 3562.64(12) and 1942.06(6)~keV.  The upper limit of the 5505~keV \emph{E5} ground-state transition intensity was determined to be 0.32\% (relative to the 3562.64(12)~keV transition, 90\% confidence limit) from the Compton background level in the $\gamma$ singles data. Transitions feeding the isomer were found by gating on the 1942.06(6)~keV and 3562.64(12)~keV stopped lines and projecting out Doppler-corrected energies of coincident $\gamma$ rays.  A total of 13 feeding transitions were observed, shown in Figures \ref{fig:32si_decayscheme} and \ref{fig:EDopp_spectra_32si} and listed in Table \ref{tab:32si_levels}.

\par
\begin{figure}
\begin{center}
\includegraphics[width=1.0\columnwidth]{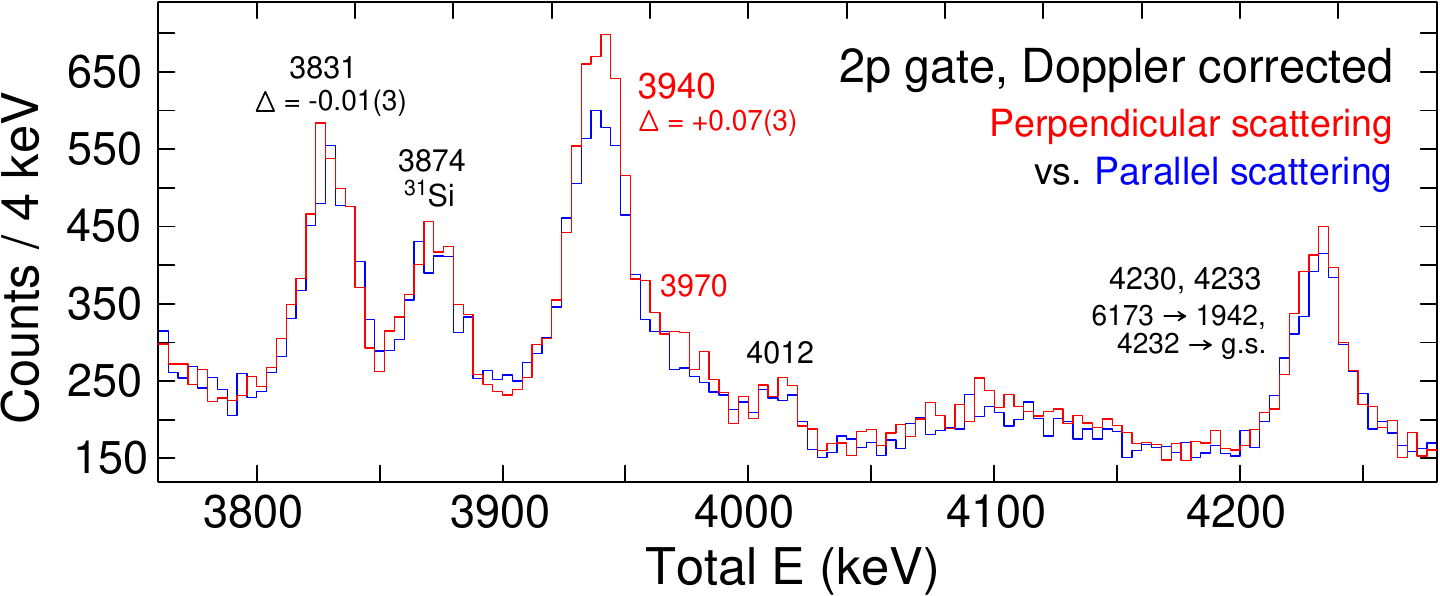}
\end{center}
\caption{Doppler-corrected $2p$ ($^{32}$Si) gated $\gamma$ singles spectrum for events which Compton scatter either in a perpendicular (red) or parallel (blue) direction with respect to the reaction plane, in the vicinity of the 3940.1(12)~keV transition.}
\label{fig:asym_32si}
\end{figure}

 {\bf 5883 keV level:} The $2^+_1$ state at 1942.12(6)~keV is fed by both the $5^-$ isomer and several shorter-lived states.  The most intense non-isomeric feeding transition is at 3940.1(12)~keV, defining a previously unobserved level at 5882.5(12)~keV.  The DCO ratio obtained for this transition when gating on the $2^+_1 \rightarrow 0^+_1$ transition was 1.00(9), indicating quadrupole character. The angular distribution coefficients $a_2 = 0.12(4)$, $a_4 = -0.06(5)$ also favor a quadrupole assignment. This transition shows a significant preference for Compton scattering perpendicular to the reaction plane (see Figure \ref{fig:asym_32si}), resulting in a $\Delta_{asym}$ value of +0.07(3) which supports an \emph{E2} assignment. The high intensity of the 3940.1(12)~keV transition relative to other lines suggests that the parent 5882.5(12)~keV level is yrast or near-yrast, so we assign it $J^{\pi} = 4^+$.  The lifetime of 18(8) fs for this level was determined simultaneously with the higher-lying 9852(2)~keV level, due to the overlapping lineshapes of the transitions depopulating these levels.  Feeding from the 7474.2(9) and 8900(3)~keV states was also considered, with the feeding intensity and lifetime varied within the $1\sigma$ confidence bounds measured for all observed feeding states to obtain the feeding-related uncertainty in the lifetime.

\par {\bf 5773, 6348 keV levels:} These levels are closely connected in the decay scheme. The 5773.1(12)~keV level was previously assigned $J = (1,2,3)$ from $^{30}$Si($t,p$) data \cite{1974Gu11}.  Our angular distribution and $R_{DCO}$ data confirms that the 3830.8(12)~keV transition depopulating this level has dipole character. This level is fed by a 575.4(3)~keV dipole transition from the 6347.7(3)~keV level, which also feeds the $J^{\pi} = 5^-$ isomer by a 842.7(3)~keV dipole transition of similar intensity. Both of these feeding transitions contained a significant stopped component in the DSAM target data, leading to a mean lifetime of 980(140) fs for the 6347.7(3)~keV level. The observed decay scheme is consistent with $J=3$ for the 5773.1(12)~keV level and $J=4$ for the 6347.7(3)~keV level.  The polarization asymmetry data was inconclusive, however as the feeding scheme involves the $J^{\pi} = 5^-$ isomer, we conclude negative parity is more likely for both levels. We assign $J^{\pi} = 3^{(-)}$ for the 5773.1(12)~keV level and $J^{\pi} = 4^{(-)}$ for the 6347.7(3)~keV level.  Transitions from these levels to the $3^-_1$ state at 5288.8(9)~keV were not observed, with branching upper limits of 1.2\% for the 
$3^{(-)} \rightarrow 3^-_1$ transition (at 90\% confidence, relative to the 3830.8(12)~keV transition) and 5.1\% for the 
$4^{(-)} \rightarrow 3^-_1$ transition (at 90\% confidence, relative to the 575.4(3)~keV transition) determined based on the Compton background. These values are consistent with shell model calculations, discussed further below.

\par {\bf 6837 keV level:} This is a newly observed level, inferred from a 1332.0(5)~keV transition observed in coincidence with the stopped lines of the isomeric cascade.  Our angular distribution and $R_{DCO}$ data indicate that the 1332.0(5)~keV transition is mixed dipole/quadrupole, restricting the spin of this level to $4 \leq J \leq 6$. Since this level decays to the $5^-$ isomer, is is likely also negative parity. There is a more probable candidate for the $6^-_1$ level observed at 7868.9(7)~keV discussed further below, so we assign $J^{\pi} = (4^-,5^-)$ to the 6837.0(5)~keV level.

\par {\bf 6850 keV level:} This is another newly observed level, based on 1562(2) and 4906(3) keV transitions to the $3^-$ level at 5288.8(9) keV and the $2^+_1$ level at 1942.06(6) keV. Our lifetime measurement of $< 120$ fs for this level suggests $|\Delta J| \leq 2$ for both transitions. The 4906(3) keV transition has mixed dipole/quadrupole character based on its angular distribution, with the stretched quadrupole case ruled out by $a_4 > 0$. We therefore assign $J = (2,3)$ for this level.

\par {\bf 7057 keV level:} The angular distribution and DCO ratio data of the 1767.7(10) keV transition depopulating this level was inconclusive due to the limited statistics.  However, the short lifetime of the parent level suggests \emph{E2} or lower multipolarity for this transition, and therefore $J \leq 5$ for the parent level.

\par {\bf 7474 keV level:} Several weak transitions were observed depopulating a level at 7474.2(9)~keV, which may correspond to a previously observed level at 7482(9)~keV from $^{30}$Si$(t,p)$ data \cite{32si_fortune}.  The decays are to states with $2 \leq J \leq 4$, with a level lifetime of 80(30) fs suggesting transition multipolarities of \emph{E2} or lower.  This restricts the 7474.2(9)~keV level to $J=2,3,4$. The $R_{DCO}$ data favours dipole assignments for the transitions from this state to lower-lying $J=4$ states.  We therefore assign $J=(3^-)$ to this state.  The negative parity assignment is preferred due to the similar branching between observed transitions --- in the positive parity case, the branching would likely be dominated by \emph{M1/E2} transitions to the lower-lying $2^+$ states.

\par {\bf 7869 keV level:} A new level at 7868.9(7)~keV was inferred from a very intense 2363.9(7)~keV transition observed in coincidence with the stopped lines of the isomeric cascade.  The angular distribution and $R_{DCO}$ data suggests a quadrupole transition, however the short DSAM lifetime of 43(9) fs rules out an M2 assignment and discourages a pure \emph{E2} transition as this would correspond to a B(E2) value of $43^{+11}_{-7}$ W.u., significantly stronger than any other transition observed in this nucleus.  The polarization asymmetry $\Delta_{asym} = -0.14(8)$ favours an \emph{M1} assignment.  Most likely this is a mixed \emph{M1/E2} transition with a significant \emph{E2} component. Assuming $\sigma/J_i \leq 0.4$ (based on the measured value for the lower 3562.64(12) keV transition), the \emph{E2/M1} mixing ratio $\delta = +1.0(6)$ derived from the angular distribution has a rather large uncertainty, with lower values of $\delta$ corresponding to lower $\sigma/J_i$ values. As a result, the \emph{E2} strength of this transition could not be well constrained, but the \emph{M1} strength is clearly quite small ($2.8^{+3.3}_{-1.5} \times 10^{-2}$~W.u.).  The strong population of this state suggests that it is yrast, which would restrict its spin to $J \geq 6$.  We therefore assign $J^{\pi} = (6)^-$ to the 7868.9(7)~keV level.  The low \emph{M1} strength of the 2363.9(7)~keV transition and the lack of an observed transition to the $4^{(-)}$ level at 6347.7(3)~keV (an upper intensity limit of 2.2\% relative to the 2363.9(7)~keV transition was obtained at 90\% confidence) are both supported by shell model calculations and discussed further below.

\par {\bf 7902, 8309 keV levels:} We have placed levels at 7901.8(10) and 8309.0(14)~keV which may correspond to the levels previously reported at 7887(18) and 8321(8)~keV in Ref.~\cite{32si_fortune}. For the 7901.8(10) level, the short lifetime of 160(80) fs and significant branching to the lower $J^{\pi} = 4^{(-)}$ level suggests $J \leq 6$.  For the 8309.0(14)~keV level, the DCO ratio of the depopulating 2808(3)~keV transition suggests a mixed dipole/quadrupole nature, consistent with the previous spin-parity assignment of $J^{\pi} = 5^-$, which we tentatively adopt.

\par
\begin{figure}
\begin{center}
\includegraphics[width=1.0\columnwidth]{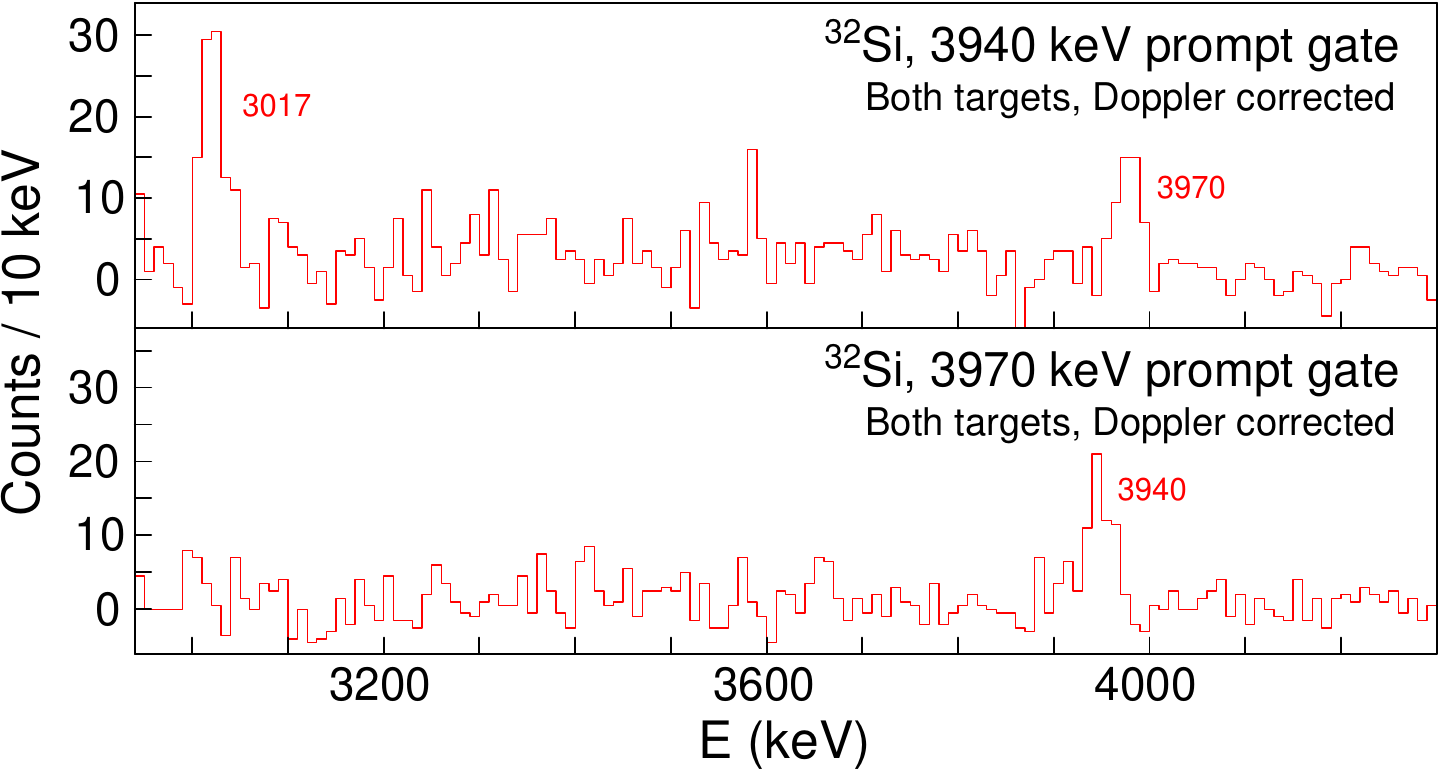}
\end{center}
\caption{Background-subtracted Doppler-corrected $\gamma$-ray spectra gated on members of the 3940.1(12)~keV and 3970(2)~keV doublet in $^{32}$Si.  All labelled transitions are newly observed.}
\label{fig:EDopp_spectra_32si_6plus_doublet}
\end{figure}

 {\bf 8900, 9852 keV levels:} Two new levels at 8900(3) and 9852(2)~keV were identified from weak transitions at 3017(3) and 3970(2)~keV observed in coincidence with the 3940.1(12)~keV $4^+ \rightarrow 2^+$ transition. Gamma-ray spectra gated on the members of the $\sim$3.95~MeV doublet are shown in Figure \ref{fig:EDopp_spectra_32si_6plus_doublet}.  The angular distribution and $R_{DCO}$ data favor stretched quadrupole assignments for both of the upper transitions, so we assign $J^{\pi} = (6^+)$ for both of these levels. The level energy systematics of neighbouring isotopes and isotones suggest that the $^{32}$Si $6^+_1$ level energy is probably in the range of 9-10~MeV. 

\par {\bf 9192, 9254 keV levels:} Two levels were placed at 9192.2(14) and 9254(2)~keV based on prompt 3687.0(14) and 3749(2)~keV transitions which feed the $5^-$ isomer.  The $R_{DCO}$ values for both transitions suggest mixed dipole/quadrupole character.  Based on this information and the lack of any other observed branching, both levels are assigned $J \leq 6$.  The negative parity case is more likely, however one or more of these states could be a 2-particle 2-hole $6^+$ state which preferentially decays to 1-particle 1-hole negative-parity states rather than a 0-particle 0-hole $4^+$ state.

\par {\bf Other levels:} High-energy transitions at 4770(4), 6702(5), and 6912(7) keV were interpreted as the decays of previously observed levels at 10275(4), 8645(5), and 8855(7) keV, respectively.  Additionally, transitions at 3189(3) and 4527(3) keV define a newly observed level at 10029(2) keV.  Limited statistics precluded direct spin-parity assignments for these levels, however $|\Delta J| \leq 2$ can be assumed for the depopulating transitions based on the lifetime measurements, resulting in the upper limits for $J$ reported in Table \ref{tab:32si_levels}.

\section{$^{29}$Al Data}
\label{sec:analysis_29al}

\begin{figure*}
\begin{center}
\includegraphics[width=2.05\columnwidth]{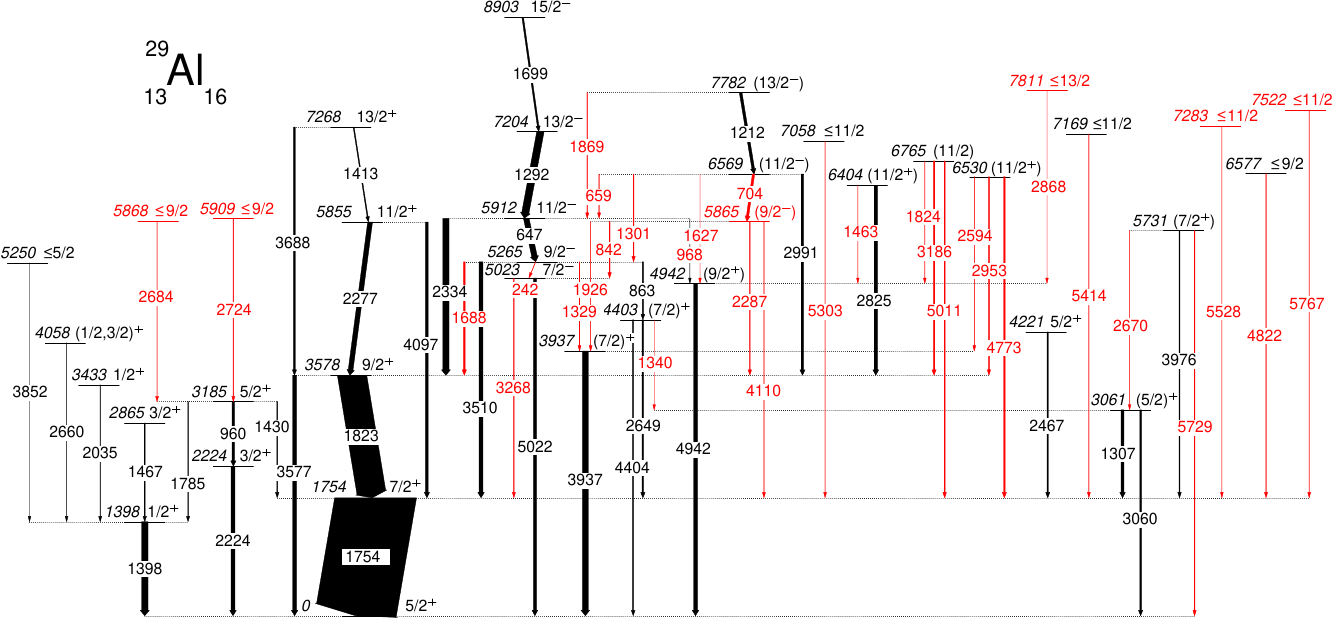}
\end{center}
\caption{Decay scheme of $^{29}$Al containing levels and transitions observed in this work.  Line widths indicate relative intensities of each transition, diagonal lines correspond to in-band \emph{M1/E2} transitions. Newly observed levels and transitions are in red. Some spin values are reported as upper limits.}
\label{fig:29al_decayscheme}
\end{figure*}

\begin{figure}
\begin{center}
\includegraphics[width=1.0\columnwidth]{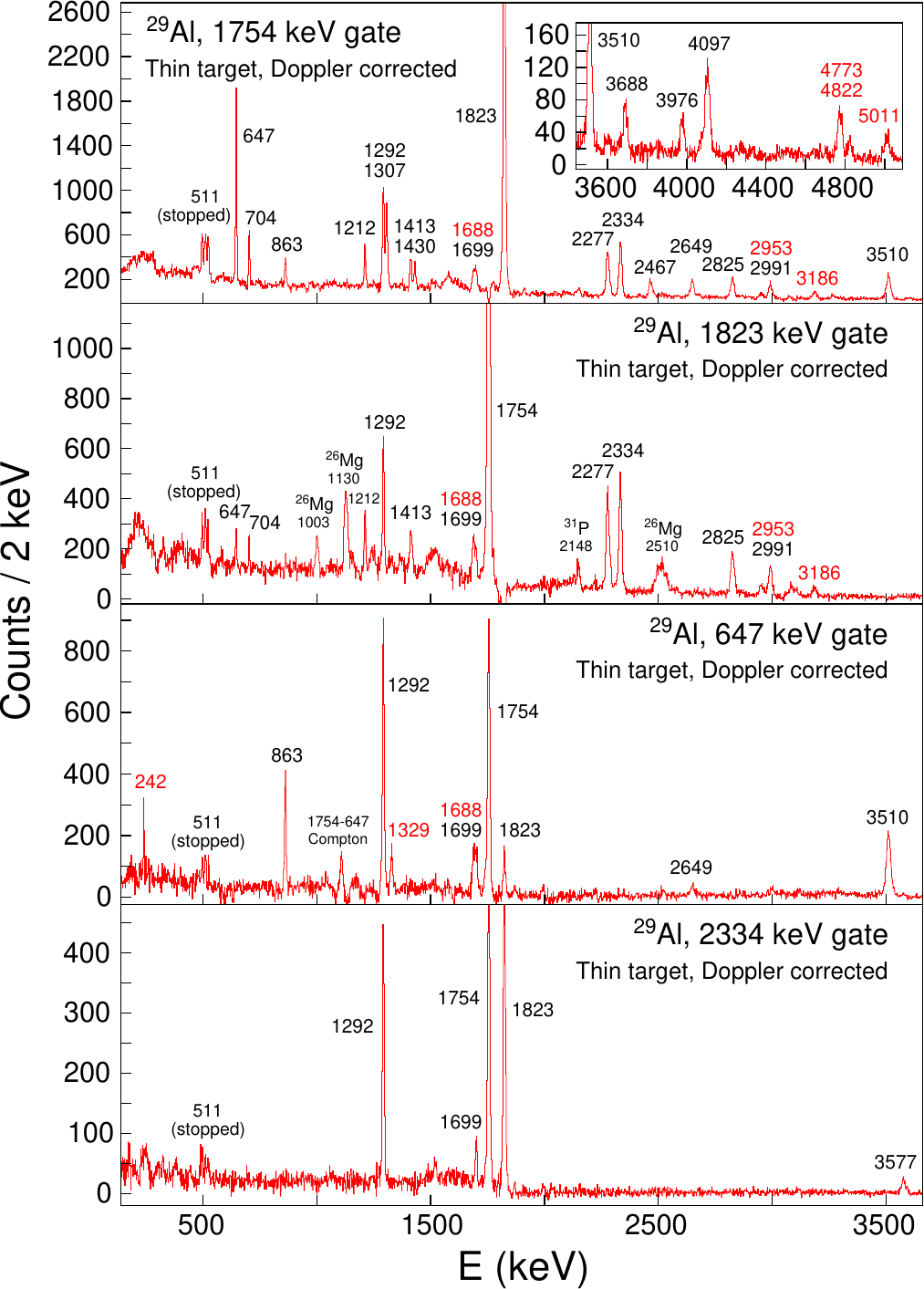}
\end{center}
\caption{Background-subtracted Doppler-corrected $\gamma$-ray spectra gated on transitions in $^{29}$Al.  Previously unobserved transitions are labelled in red.}
\label{fig:EDopp_spectra_29al}
\end{figure}

\begin{figure}
\begin{center}
\includegraphics[width=1\columnwidth]{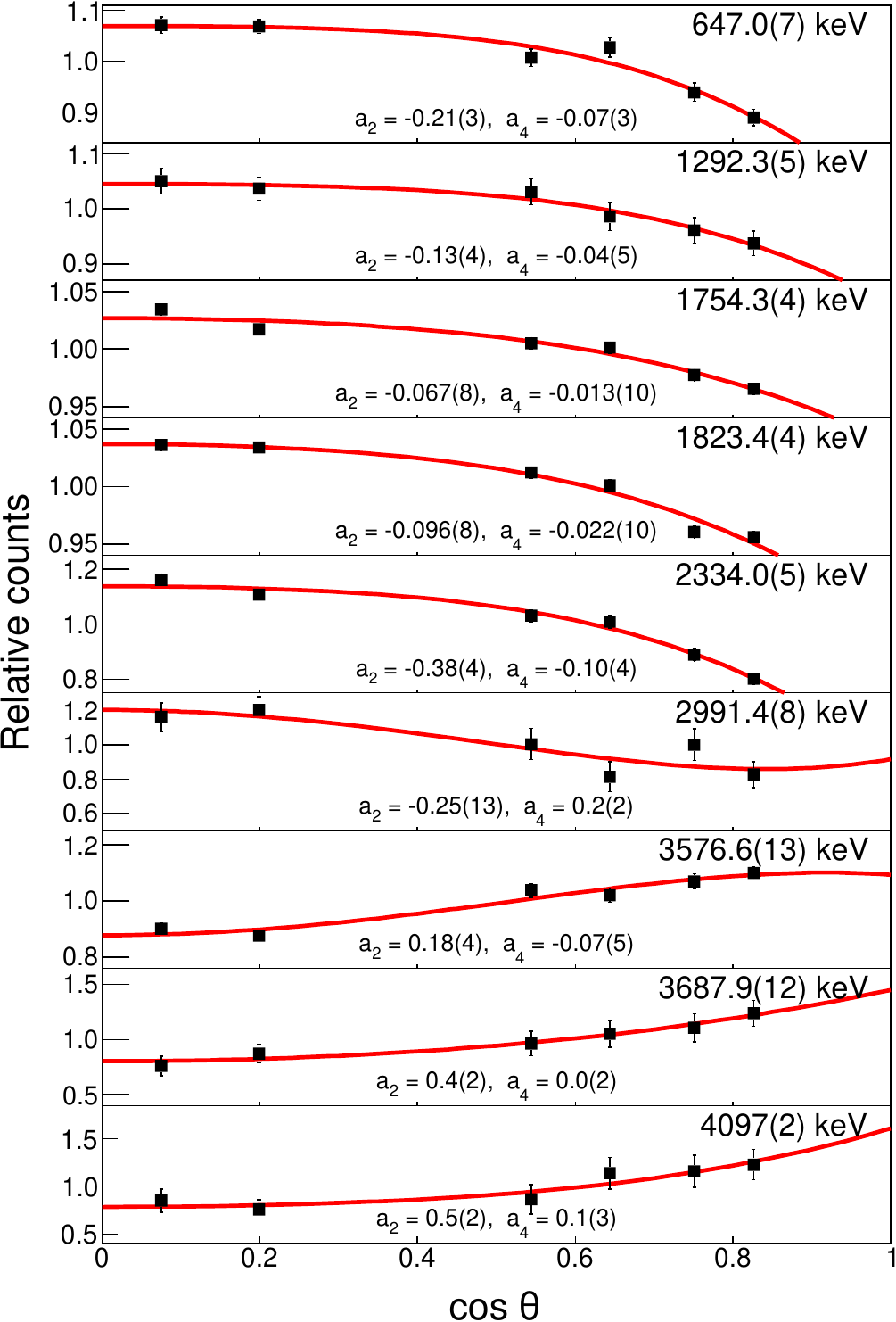}
\end{center}
\caption{Angular distributions of various transitions in $^{29}$Al, with fits to Eq. \ref{eq:angdist}.}
\label{fig:29al_angdist}
\end{figure}

\par A list of observed levels and transitions in $^{29}$Al is shown in Table \ref{tab:29al_levels}, and the corresponding level scheme in Figure \ref{fig:29al_decayscheme}.  Our level scheme is in general agreement with the previous data of Refs. \cite{dungan_2016,sultana_2018} with some minor differences (such as the placement of the 704.0(7) keV transition) and addition of several weak transitions. Doppler-corrected $\gamma$-gated spectra are shown in Figure \ref{fig:EDopp_spectra_29al}.  Angular distribution fit coefficients, $R_{DCO}$, and $\Delta_{asym}$ values are shown in Table \ref{tab:gamma_dist_table} alongside the $^{32}$Si data, and angular distribution fits are shown in Figure \ref{fig:29al_angdist}. Additional details on specific levels and transitions follows:

\par {\bf Lower spin levels:} Previously identified low spin levels at 1398.0(5) and 2224.3(4) keV were weakly populated. The only linking transition observed between the bands built on these levels and the rest of the level scheme was the 1429.8(5) keV transition between the $5/2^+$ level at 3184.7(5) keV and the strongly populated $7/2^+$ level at 1754.3(4) keV.  Two new levels were  identified at 5868.3(15) and 5909.3(13) keV based on transitions feeding the previously observed 3184.7(5) keV level.  The $\gamma$-ray angular distribution data was inconclusive for both transitions due to limited statistics. The parent levels have short lifetimes ($\tau_{mean} < 200$ fs), suggesting $\Delta J \leq 2$ for the depopulating transitions and therefore $J \leq 9/2$ for both levels. The 5909.3(13) keV level is close in energy to another level placed at 5911.9(6) keV but has a different $\gamma$-decay scheme.

\par {\bf Main band:} There is a strongly populated cascade linking the ground state to levels at 1754.3(4), 3577.7(5), 5854.8(7), and 7267.5(7) keV.  The non-crossover transitions were all consistent with mixed dipole/quadrupole character based on the $\gamma$-ray angular distribution data, and the 1754.3(4), 1823.4(4), and 2277.4(5) keV transitions were firmly assigned \emph{M1}+(\emph{E2}) multipolarity based on angular distribution and $\Delta_{asym}$ measurements.  Three crossover transitions were also observed, all with quadrupole-type angular distributions, and the 3576.6(13) keV ground-state transition was firmly assigned \emph{E2} multipolarity from its $\Delta_{asym}$ value.  The summing contributions for the crossover transitions were minimal, with $\sim$1.5\% of the observed $\gamma$ rays in the 3576.6(13) keV photopeak corresponding to summing of the intense 1754.3(4) and 1823.4(4) keV transitions, and significantly smaller summing effects for the other transitions. The DSAM lifetime measurements for the 1754.3(4) and 3577.7(5) keV levels required significant feeding correction. In both cases, the effective lifetime of the in-band feeding was approximately 50 fs (determined from the feeding-uncorrected lineshapes of the 1823.4(4) and 2277.4(5) keV transitions), which was longer than the feeding-corrected lifetime of either level. The short lifetimes of all levels in this band ($\tau_{mean} < 60$ fs) is evidence that none of the observed transitions involves a change in parity. As the lowest levels in this band were previously assigned $J^{\pi} = 5/2^+$ (ground state) and $7/2^+$, we assign $J^{\pi} = 9/2^+, 11/2^+, 13/2^+$ for the upper members of the band.

\par {\bf Negative parity band:} A strong cascade defining levels at 5265.0(7), 5911.9(6), 7204.2(7), and 8903.3(8) keV was originally reported in Ref.~\cite{dungan_2016} and interpreted in Ref.~\cite{sultana_2018} as a magnetic rotation (or shears) band based on the presence of intense in-band $M1$ transitions between levels with energies following the $E - E_0 \approx (J - J_0)^2$ energy-spin relation expected for magnetic rotation \cite{clark_shears_2000} with $J_0 = 9/2$.  A signature of magnetic rotation is $B(M1; J \rightarrow J-1)$ values which decrease with increasing $J$, however this trend could not be confirmed in Ref.~\cite{sultana_2018} as only upper limit lifetimes could be determined for the higher spin levels in this band, and \emph{M1/E2} mixing ratio measurements were unavailable. The band members were assigned $J^{\pi} = (9/2^+)$ through $(15/2^+)$ based on $J_0 = 9/2$ and a comparison to shell model calculations using the USDA interaction in the $sd$ valence space.
\par In our data, we confirm the previously reported transitions, as well as a newly observed 241.5(9) keV transition linking the band to a lower-lying 5022.6(13) keV level. Based on the established rotor-like energy spacing, we place this 5022.6(13) keV level as the band-head with $J_0 = 7/2$ (rather than the 5265.0(7) keV level with $J_0 = 9/2$ suggested in Ref.~\cite{sultana_2018}). The 5022.6(13) keV level was previously assigned $L=3$ in $^{27}$Al($t,p$) data \cite{bland_27alt_1984}, which implies negative parity for that level and by extension the entire band. An $L=3$ assignment was also given for a 5253(8) keV level which may correspond to the $J=9/2$ member of the band at 5265.0(7) keV.  In our data, the angular distribution and Compton polarization data of Table \ref{tab:gamma_dist_table} favours \emph{M1}(+\emph{E2}) assignments for the strong 647.0(7) and 1292.3(5) keV transitions within the band, and (\emph{E1}) assignments for the strong 2334.0(5) and 3510.0(10) keV transitions out of the band, all consistent with negative parity assignments for the band members.  The negative parity case is also supported by the non-observation of any transitions which would be \emph{E2} if the band is positive parity, but \emph{M2} if the band is negative parity. For example, no $\Delta J = 2$ transitions from the 5265.0(7) or 5911.9(6) keV levels to the ground-state band were observed, whereas multiple intense $\Delta J = 2$ crossover transitions were observed within the ground-state band itself.  We therefore assign the members of the excited rotor-like band $J^{\pi} = 7/2^-$, $9/2^-$, $11/2^-$, $13/2^-$, and $15/2^-$.
\par A search for higher spin members of this band and in-band crossover transitions turned up negative.  Branching fraction limits for \emph{E2} crossover transitions were determined from the Compton background level: $< 5.2$\% ($15/2^- \rightarrow 11/2^-$), $< 4.5$\% ($13/2^- \rightarrow 9/2^-$), $< 1.2$\% ($11/2^- \rightarrow 7/2^-$), all reported to 90\% confidence.

\par {\bf 3937 keV level:} The level at 3937.1(14) keV was previously assigned $J^{\pi} = (3/2,7/2)^+$ in Ref.~\cite{bland_27alt_1984}.  Based on the strong population of this level in our fusion-evaporation data and a comparison of the intensity of the 3936.8(14) keV line to transitions depopulating other known $3/2^+$ levels at 2224.3(4) and 2865.3(7) keV, we conclude that the higher spin assignment for the 3937.1(14) keV level is significantly more likely and assign $J^\pi = (7/2)^+$.

\par {\bf 4403 keV level:} This level was previously assigned $J^{\pi} = (7/2, 9/2)^+$ in Ref.~\cite{bland_27alt_1984}. Our angular distribution data favours a $\Delta J = 1$, D+Q assignment for the 4403.7(14) keV ground-state transition (from $a_4 > 0$) and therefore a $J^\pi = (7/2)^+$ assignment for this level.

\par {\bf 4942 keV level:} The 4942.0(17) keV level did not have any previous spin-parity assignment.  Our angular distribution data favours a quadrupole assignment for the 4941.6(17) ground-state transition (from $a_2 > 0$), and the very short lifetime of the parent level rules out an \emph{M2} assignment (since the $B(M2)$ value would be $\geq 73$ W.u., much higher than the recommended upper limit of 5 W.u. \cite{endt_strengths_1993}).  A mixed \emph{M1/E2} assignment for this transition cannot be ruled out from the angular distribution, however this would result in an \emph{M2} assignment for the 968.1(8) keV feeding transition, with an unrealistically large transition strength of $\sim$300~W.u.  We therefore assign $J^\pi = (9/2)^+$ for the 4942.0(17) keV level.

\par {\bf 5731 keV level:} We assigned $J^{\pi} = (7/2^+)$ for a previously observed level at 5730.7(10) keV, based on the angular distributions of the 3976.1(15) and 5729(2) keV transitions depopulating this state, which favour $\Delta J = 0,2$ and $\Delta J = 1$ assignments, respectively.  The short lifetime of the parent level ($< 14$ fs) suggests that all depopulating transitions have \emph{M1} and/or \emph{E2} multipolarity.

\par {\bf 5865, 6569, 7782 keV levels:} These levels are linked by strong 704.0(7) and 1212.3(5) transitions which were both firmly assigned \emph{M1/E2} multipolarity based on the angular distribution and $\Delta_{asym}$ measurements.  Both of these transitions were previously reported in Ref.~\cite{dungan_2016}, however our placement of the 704.0(7) keV is different since we observed it in coincidence with a new 2286.9(8) keV transition rather than the previously reported 2277.4(5) keV transition.  This reassignment, along with newly observed transitions at 841.7(8), 1926.3(8), and 4110.1(15) keV, defines a new level at 5864.5(8) keV. The angular distribution of the 4110.1(15) keV transition favors a mixed dipole/quadrupole assignment (from $a_2 < 0$, $a_4 \geq 0$), which when combined with the \emph{M1/E2} in-band transitions suggests $J = 9/2, 11/2, 13/2$ for this band. This is further supported by a mixed dipole/quadrupole assignment for the 2991.4(8) keV transition out of the band.  This band is probably negative parity, since there are several transitions linking it to the strongly populated negative parity band built on the 5022.6(13) keV state.  The decay scheme of the 5864.5(8) keV level also suggests negative parity, as the branching of the 841.7(8) keV transition to a $7/2^-$ state is similar to that of the 4110.1(15) keV transition to a $7/2^+$ state, whereas the higher energy transition would dominate if the configurations of the parent and daughter states were similar.  Finally, the $\Delta_{aysm}$ measurement of the 2991.4(8) keV transition slightly favours an \emph{E1/M2} assignment, which would be consistent with negative parity for the parent 6569.2(7) keV level.  We assign $J^{\pi} = (9/2^-)$, $(11/2^-)$, and $(13/2^-)$ for the members of this band.

\par {\bf 6404 keV level:} This level was assigned $J^{\pi} = (11/2^+)$ based on the dipole/quadrupole-type angular distribution of the 2825.4(7) keV transition.  Positive parity is assumed based on the very short lifetime of the level ($< 5$ fs).

\par {\bf 6530 keV level:} The angular distribution data for transitions depopulating this level is mostly inconclusive, however a quadrupole assignment is slightly preferred for the 4773(2) keV transition (from $a_2 > 0$).  Combined with the short lifetime of the parent level ($< 30$ fs), this supports a $J^{\pi} = (11/2^+)$ assignment. 

\par {\bf 6577 keV level:} The energy of this level is consistent with a level previously observed in Ref.~\cite{bland_27alt_1984}.  The 4822(2) keV transition depopulating this level was assigned mixed dipole/quadrupole character from the angular distribution data, implying $J \leq 9/2$ for this level.

\par {\bf 6765 keV level:} The level energy is consistent with a level previously reported in Ref.~\cite{bland_27alt_1984}, however the transitions depopulating this level were not previously observed. The angular distribution data for the 5011(2) keV transition to the first $7/2^+$ level favors a quadrupole assignment (from $a_2 > 0$), while a dipole assignment is slightly favored for the 3186.4(10) keV transition to the first $9/2^+$ level.  The $\Delta_{asym}$ value of the 3186.4(10) keV transition slightly favours an \emph{E1} assignment, suggesting negative parity for the parent level, but the low statistics prevented a definite assignment.  We assign $J=(11/2)$ for this level.

\par {\bf Other levels:} High-energy 5303(4), 5414(2), 5528(3), 5767(3), and 2868(2) keV transitions were interpreted as the decays of levels at 7058(4), 7169(2), 7283(3), 7522(3), and 7811(2) keV, where the first two are consistent with previously observed levels and the latter three are newly observed.  Limited statistics for these transitions precluded direct spin-parity assignments, however the short lifetimes of all of these levels suggests that the corresponding transitions are of \emph{E2} or lower multipolarity, implying $J \leq 13/2$ for the 7811(2) keV level and $J \leq 11/2$ for the other levels.

\begin{longtable}[]{llllr}
    \caption{List of $^{29}$Al levels and $\gamma$ rays observed in this work.  Items in {\bf bold} are newly observed or measured ($I_{\gamma}$ values resulting in new branching ratios are also highlighted). Lifetime limits are reported to 90\% confidence, all other quoted uncertainties are at 1$\sigma$ confidence.}\\
    \hline \hline
    \rule{0pt}{2.5ex}$E_{level}$ (keV) & $E_{\gamma}$ (keV) & $I_{\gamma, rel}$ & $\tau_{mean}$ (fs) & $J^{\pi}$ \\ \hline
    \endfirsthead
    
    \caption{$(Continued.)$}\\
    \hline \hline
    \rule{0pt}{2.5ex}$E_{level}$ (keV) & $E_{\gamma}$ (keV) & $I_{\gamma, rel}$ & $\tau_{mean}$ (fs) & $J^{\pi}$ \\ \hline
    
    \endhead
    \hline
    \endfoot
    \hline \hline
    \endlastfoot
       
\rule{0pt}{2.5ex}1398.0(5)   & 1398.0(5)  & 8.1(3) & $> 2500$ & $1/2^+$ \\
                 1754.3(4)   & 1754.3(4)  & 100.0(2) & 17(2) & $7/2^+$ \\
                 2224.3(4)   & 2224.2(4)  & 4.9(12) & 41(10) & $3/2^+$ \\
                 2865.3(7)   & 1467.2(5)  & 0.99(6) & 115(13)$^{\dag}$ & $3/2^+$ \\
                 3061.3(6)   & 1307.0(5)  & 4.1(2) & 31(5)$^{\dag}$ & $(5/2)^+$ \\
                             & 3060(2)    & 2.2(8) & & \\
                 3184.7(5)   & 959.8(6)   & 3.35(6) & 131(7) & $5/2^+$ \\
                             & 1429.8(5)  & 1.23(8) & & \\
                             & 1785.2(4)  & 0.70(7) & & \\
                 3433.1(10)  & 2035.0(9)  & 0.48(6) & $\mathbf{220(40)^{\dag}}$ & $1/2^+$ \\
                 3577.7(5)   & 1823.4(4)  & 36.1(5) & 11(9) & $\mathbf{9/2^+}$ \\
                             & 3576.6(13) & 4.75(9) & & \\
                 3937.1(14)  & 3936.8(14) & 7.3(2) & 80(30) & $\mathbf{(7/2)^+}$ \\
                 4058.1(13)  & 2659.9(12) & 0.24(4) & $< 26$ & $(1/2,3/2)^+$ \\
                 4221.1(7)   & 2466.6(6)  & 2.06(12) & $22(12)^{\dag}$ & $5/2^+$ \\
                 4403.4(7)   & \textbf{1340.1(12)}  & \textbf{0.15(3)} & $19(8)$ & $(7/2)^+$ \\
                             & 2648.6(7)  & 2.33(14) & & \\
                             & 4403.7(14)    & 1.5(4)   & & \\
                 4942.0(17)  & 4941.6(17) & 4.73(9) & 15(7)$^{\dag}$ & $\mathbf{(9/2)^+}$\\
                 
                 5022.6(13)  & \textbf{3268(2)} & \textbf{0.68(7)} & $\mathbf{40(30)}$ & $\mathbf{7/2^-}$ \\
                             & 5022(2)    & \textbf{3.64(8)} & & \\
                 5250(2)     & 3852(2)   & 0.10(3)  & $< 140$ & $\mathbf{\leq 5/2}$ \\
                 
                 5265.0(7)   & \textbf{241.5(9)} & \textbf{0.32(6)} & 200(30) & $\mathbf{9/2^-}$ \\
                             & 862.8(7)          & \textbf{1.42(12)} & & \\
                             & \textbf{1328.5(6)} & \textbf{0.35(5)} & & \\
                             & \textbf{1687.6(8)} & \textbf{2.14(18)} & & \\
                             & 3510.0(10) & \textbf{5.3(3)} & & \\
                 5730.7(10)  & \textbf{2670(2)} & \textbf{0.16(2)} & $\mathbf{< 14}$ & $\mathbf{(7/2^+)}$ \\
                             & 3976.1(15)          & \textbf{1.17(10)} & & \\
                             & \textbf{5729(2)}    & \textbf{0.79(3)} & & \\
                             
                 5854.8(7)   & 2277.4(5)  & 5.9(5)   & 50(5)$^{\dag}$ & $\mathbf{11/2^+}$ \\
                             & 4097(2) & 3.6(2) & & \\
         \textbf{5864.5(8)}  & \textbf{841.7(8)}   & \textbf{1.1(4)} & $\mathbf{80(30)}$ & $\mathbf{(9/2^-)}$ \\
                             & \textbf{1926.3(8)}  & \textbf{0.40(5)} & & \\
                             & \textbf{2286.9(8)}  & \textbf{1.0(3)} & & \\
                             & \textbf{4110.1(15)} & \textbf{1.3(5)} & & \\
         \textbf{5868.3(15)} & \textbf{2683.5(13)} & 0.34(9) & $\mathbf{130(50)^{\dag}}$ & $\mathbf{\leq 9/2}$ \\
         \textbf{5909.3(13)} & \textbf{2724.3(11)}  & 0.51(9) & $\mathbf{75(19)^{\dag}}$ & $\mathbf{\leq 9/2}$ \\
                 5911.9(6)   & 647.0(7)            & 6.81(12) & 100(15) & $\mathbf{11/2^-}$ \\
                             & \textbf{968.1(8)}  & \textbf{0.080(2)} & & \\
                             & 2334.0(5)           & 7.83(13) & & \\
                 6403.6(8)   & \textbf{1462.7(14)} & \textbf{0.15(3)} & $\mathbf{< 5}$ & $\mathbf{(11/2^+)}$ \\
                             & 2825.4(7)           & \textbf{4.0(3)}  & & \\
                 6529.7(11)  & \textbf{2593.8(12)} & \textbf{0.18(3)}  & $\mathbf{< 30}$ & $\mathbf{(11/2^+)}$ \\
                             & \textbf{2952.6(13)} & \textbf{0.93(11)} & & \\
                             & \textbf{4773(2)} & \textbf{2.20(14)} & & \\
                 6569.2(7)   & \textbf{658.5(9)}   & \textbf{0.5(2)} & $\mathbf{< 50}$ & $\mathbf{(11/2^-)}$ \\                 
                             & \textbf{704.0(7)}   & \textbf{2.85(6)} & & \\
                             & \textbf{1301.4(15)} & \textbf{0.9(4)} & & \\                             
                             & \textbf{1626.8(15)} & \textbf{0.13(3)} & & \\
                             & 2991.4(8)           & \textbf{3.1(2)}  & & \\
                 6577(2)     & \textbf{4822(2)}    & 0.66(7) & $\mathbf{< 230}$ & $\mathbf{\leq 9/2}$ \\
                 6765.2(9)   & \textbf{1824.3(12)} & \textbf{0.20(4)}  & $\mathbf{< 75}$ & $\mathbf{(11/2)}$ \\
                             & \textbf{3186.4(10)} & \textbf{1.87(16)} & & \\
                             & \textbf{5011(2)}    & \textbf{1.22(9)}  & & \\
                 7058(4)     & \textbf{5303(4)}    & 0.28(5) & $\mathbf{< 65}$ & $\mathbf{\leq 11/2}$ \\
                 7169(2)     & \textbf{5414(2)}    & 0.40(6) & $\mathbf{90(50)^{\dag}}$ & $\mathbf{\leq 11/2}$ \\
                 7204.2(7)   & 1292.3(5)           & 8.92(15) & $\mathbf{13(4)^{\dag}}$ & $\mathbf{13/2^-}$ \\
                 7267.5(7)   & 1412.7(5)           & 1.24(16) & $\mathbf{< 60}$ & $\mathbf{13/2^+}$ \\
                             & 3687.9(12)          & 1.98(17) & & \\
         \textbf{7283(3)}    & \textbf{5528(3)} & 0.21(5) & $\mathbf{< 45}$ & $\mathbf{\leq 11/2}$ \\         
         \textbf{7522(3)}    & \textbf{5767(3)} & 0.26(5) & $\mathbf{< 210}$ & $\mathbf{\leq 11/2}$ \\  
                 7781.5(7)   & 1212.3(5)           & \textbf{3.27(7)} & $\mathbf{30(9)^{\dag}}$ & $\mathbf{(13/2^-)}$ \\
                             & \textbf{1869.3(7)} & \textbf{0.58(10)} & & \\
         \textbf{7811(2)}    & \textbf{2868(2)} & 0.16(3) & $\mathbf{< 125}$ & $\mathbf{\leq 13/2}$ \\
                 8903.3(8)   & 1699.0(5) & 1.96(19) & $< 7$ & $\mathbf{15/2^-}$
\label{tab:29al_levels}
\end{longtable}
\begin{center}
{\footnotesize \rule{0pt}{2.5ex}$^{\dag}$Effective lifetime without any feeding correction.}
\end{center}

\section{$^{32}$Si discussion}
\label{sec:discussion_32si}

\par The intermediate-energy levels of $^{32}$Si are clearly influenced by the reduction in the $N=20$ shell gap with decreasing proton number.  In comparison to its isotone neighbour $^{34}$S, the higher spin negative-parity states of $^{32}$Si are lowered in energy relative to positive-parity states of similar spin.  This leads to the isomerism of the $5^-_1$ level, since it is close in energy to the $3^-_1$ level and lower in energy than the $4^+_1$ level, causing it to preferentially decay by a high-energy \emph{E3} transition to the $2^+_1$ level. Spin-trap isomers are fairly common in higher-mass nuclides with intruder orbitals near the Fermi energy, but this is the first case reported in an even-even $sd$ shell nuclide.

\par
\begin{figure}
\begin{center}
\includegraphics[width=1.0\columnwidth]{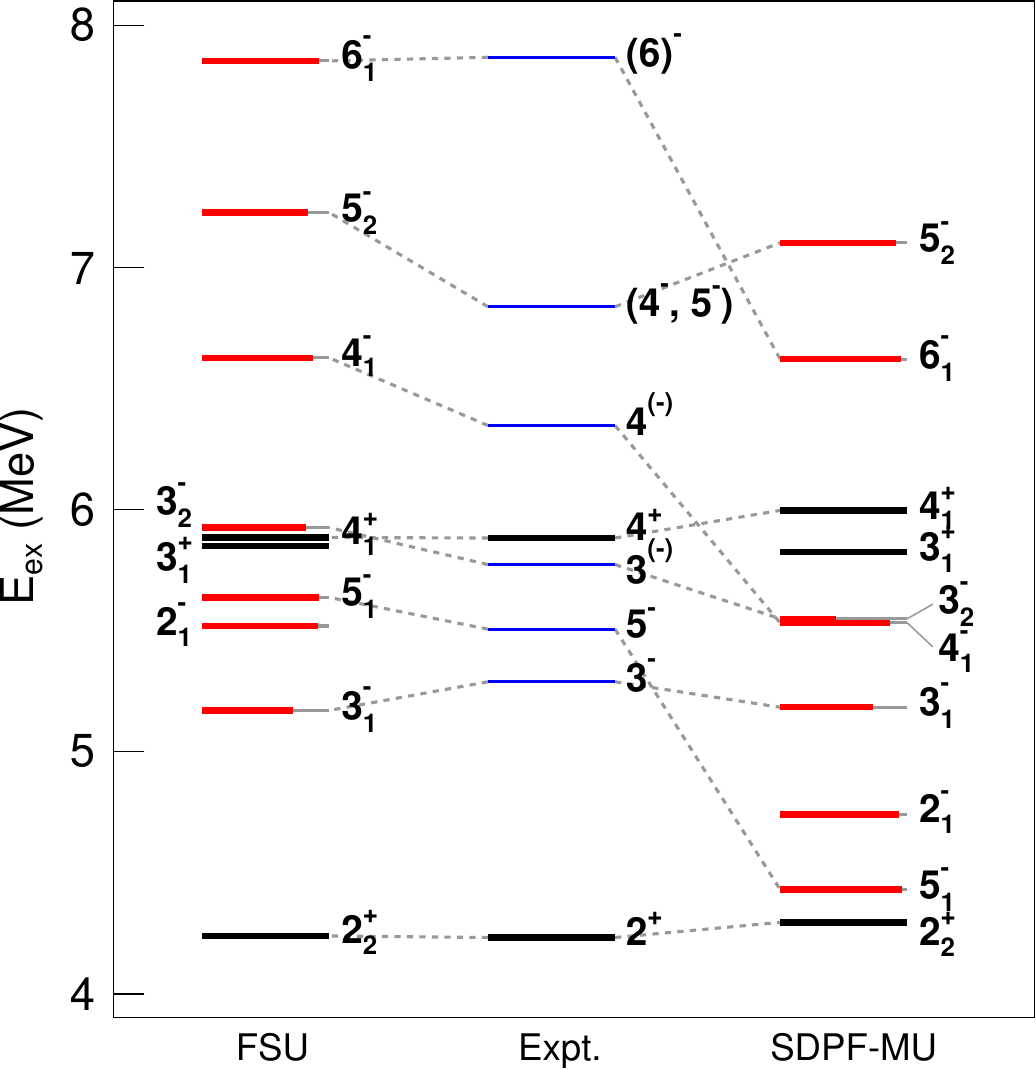}
\end{center}
\caption{Comparison of yrast and near-yrast level energies in $^{32}$Si to shell model calculations using the FSU and SDPF-MU interactions.  For calculated negative-parity states, red bars indicate the $\nu 0f_{7/2}$ orbital population.}
\label{fig:sm_levels_32si}
\end{figure}

 Shell model calculations were performed using the FSU interaction \cite{lubna2020evolution} in the $psdpf$ valence space, and the SDPF-MU interaction \cite{sdpfmu} in the $sdpf$ valence space.  With both models, negative parity $1p1h$ (1 particle --- 1 hole) states were restricted to single neutron excitation.  For the FSU model calculations, positive-parity $0p0h$ states were restricted to the $sd$ shell, giving results identical to the USDB interaction \cite{usdb} for those states.  A comparison of level energies is shown in Figure \ref{fig:sm_levels_32si}.  In general, the energies of positive-parity states are similar between both models and consistent with experiment.  For negative-parity states, the FSU model is in general agreement with the experimental results, whereas the SDPF-MU model under-predicts the energies of several states by $\sim$1~MeV, including the $5^-_1$ level.  The good agreement of the FSU calculations is unsurprising in this context, since the two-body matrix elements of the FSU interaction are fitted using similar data from this mass region --- in particular high-spin states containing excitation to the $0f_{7/2}$ orbital \cite{lubna2020evolution}.

\par Our value of $780(120)$ fs for the $2^+_1$ mean lifetime is somewhat shorter than previous values of $1300(400)$ fs and 1030$^{+170}_{-130}$ fs derived from Coulomb excitation (Coulex) \cite{Ibbotson32Si,heery_suppressed_2024}, but is more consistent with both a previous DSAM value of 900(300) fs \cite{Pronko32Si} and our shell model calculations shown in Table \ref{tab:be2_table}. The quadrupole deformation parameter $\beta_2=0.30(3)$ derived from our data is also slightly larger than the values of 0.23(4) and 0.26(2) derived from the Coulex data. When combined with existing data from this mass region, our $\beta_2$ value suggests a gradual onset of ground-state deformation along the $N=18$ isotonic chain approaching the Mg isotopes, and a sharp change along the $Z=14$ isotopic chain between $^{32}$Si (weakly deformed ground state) and $^{34}$Si (spherical), showing that persistence of the $N=20$ shell closure and that the Si isotopes are outside of the `island of inversion'. The reason for the discrepancy between Coulex and DSAM measurements is unclear, but it is probably not due to uncorrected feeding of the $2^+_1$ level, since this would result in larger $\tau_{mean}$ values for the DSAM measurements.

\par The 46.9(5) ns lifetime of the $5^-_1$ isomer implies that the transitions to the lower $2^+_1$ and $3^-_1$ levels are significantly hindered compared to the single particle values ($B(\lambda L) << 1$ W.u). As Table \ref{tab:be2_table} shows, our shell model calculations generally agree with the measured strengths of these transitions. The low transition strengths were interpreted as resulting from the differing wavefunctions of the levels involved; as shown in Figure \ref{fig:sm_levels_32si}, the calculated $\nu 0f_{7/2}$ occupancy is significantly higher for the $5^-_1$ level than the $3^-_1$ level (the $2^+_1$ level was calculated assuming no $\nu 0f_{7/2}$ occupancy), with the $5^-_1$ level dominated by the $(\nu 0d_{3/2})^{-1} \otimes (\nu 0f_{7/2})^{1}$ configuration.

\begin{table}
\caption{Transition strength values measured in the present work, compared to model calculations.  Effective charges $e_p = 1.36e$ and $e_n = 0.45e$ were used with all models.}
\centering
\begin{ruledtabular}
\begin{tabular}{lllll}
\rule{0pt}{2.5ex}\multirow{2}{*}{$J^{\pi}_i \rightarrow J^{\pi}_f$} & \multirow{2}{*}{$\lambda L$}  & \multicolumn{3}{c}{$B(\lambda L)$ (W.u.)} \\ \cline{3-5}
\rule{0pt}{2.5ex} & & Expt. & FSU & SDPF-MU \\ \hline
\rule{0pt}{2.5ex}$2^+_1 \rightarrow 0^+_1$ & \emph{E2} & 6.3$^{+1.1}_{-0.8}$ & 7.3 & 6.3 \\
\rule{0pt}{2.5ex}$4^+_1 \rightarrow 2^+_1$ & \emph{E2} & 8$^{+7}_{-3}$ & 11.2 & 8.2 \\
\rule{0pt}{2.5ex}$5^-_1 \rightarrow 3^-_1$ & \emph{E2} & $< 0.053^{*}$ & 0.08 & 0.6 \\ 
\rule{0pt}{2.5ex}$5^-_1 \rightarrow 2^+_1$ & \emph{E3} & 0.0841(10) & 0.150 & 0.012 \\ 
\rule{0pt}{2.5ex}$6^-_1 \rightarrow 4^-_1$ & \emph{E2} & $< 6.57^{*}$ & 3.3 & 2.2 \\ 
\end{tabular}
\end{ruledtabular}
\label{tab:be2_table}
{\footnotesize \rule{0pt}{2.5ex}$^*$From intensity limit of unobserved transition.}
\end{table}

\par Above the isomer, the $(6)^-$ level observed at 7868.9(7)~keV aligns well with the $6^-_1$ state predicted at 7853~keV in the FSU model calculations.  The predicted $B(M1; 6^-_1 \rightarrow 5^-_1)$ value of 0.037 W.u.~also matches the measured value of $0.028^{+0.021}_{-0.013}$~W.u for the 2363.9(7)~keV transition. The predicted $B(E2; 6^-_1 \rightarrow 4^-_1)$ value of 3.28 W.u.~suggests that the expected branching to the $4^{(-)}$ state at 6347.7(3) should be $< 1$\% of the intensity of the observed transition to the isomeric $5^-$ state, which is consistent with the upper limit of 2.2\% from the data.  The alternative spin $J=7$ spin hypothesis for the 7868.9(7)~keV level is not supported by the FSU calculations, which predict the $7^-_1$ level at $\sim$10~MeV with a $B(E2; 7^-_1 \rightarrow 5^-_1)$ value of 4.7 W.u.

\par Two candidates for the $6^+_1$ level were identified at 8900(3) and 9852(2) keV.  At high excitation energy, both $0p0h$ and $2p2h$ configurations can contribute to positive-parity states.  The FSU calculations predict the $6^+_1$ state at 9262 keV with a $2p2h$ configuration, which likely corresponds to one of the two observed levels.

\par Finally, the absence of some transitions in the level scheme is interesting.  For instance, the $4^{(-)}$ state at 6347.7(3)~keV decays to the $3^{(-)}$ state at 5773.1(12)~keV, but no transition to the lower $3^-$ state at 5288.8(9)~keV was observed. This behaviour is reproduced by the FSU model calculations assuming that the 6347.7(3)~keV level is the $4^-_1$ state.  The calculated B(M1; $4^-_1 \rightarrow 3^-_2$) value of 0.095 W.u.~is 20 times larger than the calculated B(M1; $4^-_1 \rightarrow 3^-_1$) value of 0.004 W.u, due to the higher $\nu 0f_{7/2}$ occupancy expected for the $4^-_1$ and $3^-_2$ states (0.87 and 0.82) compared to the $3^-_1$ state (0.72). From the intensity limit of the unobserved transition, B(M1; $4^-_1 \rightarrow 3^-_1$) $<$ 0.0014 W.u.~(to 90\% confidence), in reasonable agreement with the calculated values.  The same reasoning can be used to explain the non-observation of a $3^{(-)} \rightarrow 3^-$ transition between the 5773.1(12) and 5288.8(9)~keV levels.  The FSU calculations predict a B(M1; $3^-_2 \rightarrow 3^-_1$) value of 0.022 W.u.~and a B(E2; $3^-_2 \rightarrow 3^-_1$) value of 3.74 W.u.  Using the measured lifetime of $40(30)$ fs for the 5773.1(12) keV level, this would require a $3^{(-)} \rightarrow 3^-$ branching of $< 0.75$\% (and $\delta_{E2/M1} \approx 0.1$), consistent with our experimentally determined upper limit of 1.2\%.

\section{$^{29}$Al discussion}
\label{sec:discussion_29al}

\begin{figure}
\begin{center}
\includegraphics[width=1.0\columnwidth]{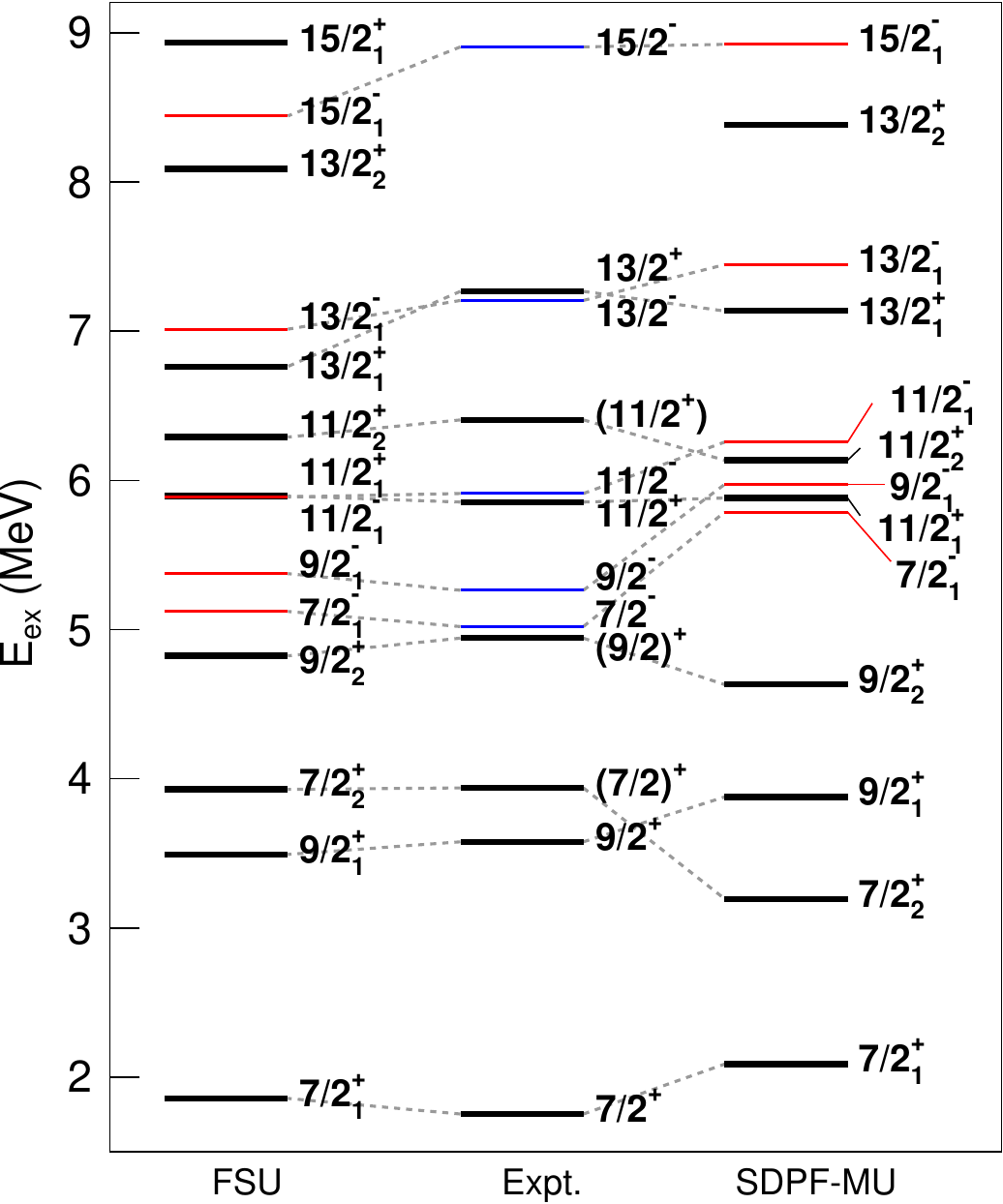}
\end{center}
\caption{Comparison of yrast and near-yrast level energies in $^{29}$Al to shell model calculations using the FSU and SDPF-MU interactions.}
\label{fig:sm_levels_29al}
\end{figure}

\par
\begin{figure}
\begin{center}
\includegraphics[width=1.0\columnwidth]{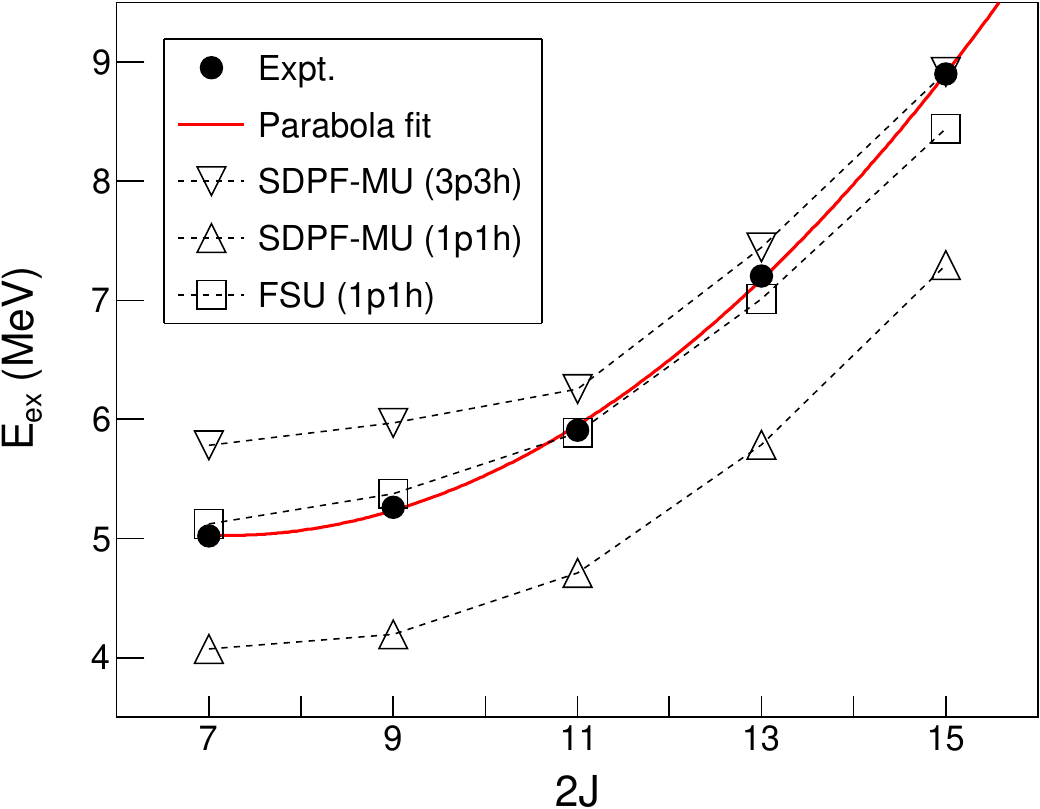}
\end{center}
\caption{Comparison of observed level energies in the intruder rotor-like band of $^{29}$Al to shell model calculations using the FSU and SDPF-MU interactions (with $sd-pf$ cross-shell truncation indicated).}
\label{fig:intruderband_29al}
\end{figure}

 Although $^{29}$Al is situated near the middle of the $sd$ shell, intermediate-energy negative-parity states play a significant role in its level scheme. A comparison of level energies to shell model calculations using the FSU (in the $psdpf$ valence space) and SDPF-MU interactions is shown in Figure \ref{fig:sm_levels_29al}.  The negative-parity states in the FSU calculations were restricted to $1p1h$ cross-shell excitation, whereas a $3p3h$ truncation was used for the SDPF-MU calculations.  We found that the SDPF-MU interaction with a $1p1h$ truncation was unable to reproduce level energies in the rotor-like negative parity band, as shown in Figure \ref{fig:intruderband_29al}.  Above 5 MeV excitation energy, these calculations predict that the yrast states switch from positive to negative parity with increasing spin, consistent with our experimental data.

\par The level energies of the main negative parity band exhibit rotor-like spacing, as discussed in Ref.~\cite{sultana_2018}.  Figure \ref{fig:intruderband_29al} shows a comparison to the shell model calculations, which generally agree with both the level spacing and spin-parity assignments determined from the data.  This level spacing could be the result of a deformed rotational band, or a magnetic rotational band (as proposed in Ref.~\cite{sultana_2018}, sometimes referred to as a shears band).  In the former case, large in-band $B(E2; J \rightarrow J-2)$ values are expected, whereas in a magnetic rotational band the $B(M1; J \rightarrow J-1)$ values are expected to decrease and $\delta_{E2/M1}$ mixing ratios are expected to increase with increasing $J$ \cite{clark_shears_2000}. Our $\delta_{E2/M1}$ measurements and derived $B(M1; J \rightarrow J-1)$ values for the 647.0(7), 1292.3(5), and 1699.0(5) keV \emph{M1/E2} transitions in this band are reported in Table \ref{tab:mixing_ratio_table}, and a comparison to shell model calculations is shown in Table \ref{tab:b_table_29Al}.  Our $B(M1; 11/2^- \rightarrow 9/2^-)$ value disagrees with the value of Ref.~\cite{sultana_2018} by a factor of 2 when taking the branching of the 647.0(7) keV transition into consideration.  The $B(M1; J \rightarrow J-1)$ values appear to increase with $J$ and the $\delta_{E2/M1}$ values do not consistently increase with $J$, both inconsistent with the expected behaviour for magnetic rotation. However, the data also seems inconsistent with a deformed rotor, based on the non-observation of crossover $\Delta J = 2$ transitions. From the measured intensity limits, the in-band $B(M1)/B(E2)$ ratios are greater than 30 $(\mu_N/eb)^2$, which is consistent with $B(M1)/B(E2) \geq 20$ $(\mu_N/eb)^2$ typically seen in shears bands \cite{clark_shears_2000}. The shell model calculations, which successfully reproduced the energy spacing for this band, also predict small $B(E2; J \rightarrow J-2)$ values for all in-band transitions, but do not reproduce the observed $B(M1)$ trend (see Table \ref{tab:b_table_29Al}).  Overall, our data shows a clear enhancement of \emph{M1} transitions relative to crossover \emph{E2} transitions in this band, but does not display the other properties typically associated with magnetic rotation.

\begin{table*}
\caption{Transition strengths measured in the negative parity band of $^{29}$Al, compared to FSU ($1p1h$ truncation, $e_p = 1.36e$, $e_n = 0.45e$) and SDPF-MU ($3p3h$ truncation, $e_p = 1.35e$, $e_n = 0.35e$) calculations.  All limits are reported to 90\% confidence.}
\centering
\begin{ruledtabular}
\begin{tabular}{lllllll}
\rule{0pt}{2.5ex}\multirow{2}{*}{$J^{\pi}_i \rightarrow J^{\pi}_f$} & \multicolumn{3}{c}{$B(M1)$ (W.u.)} & \multicolumn{3}{c}{$B(E2)$ (W.u.)} \\ \cline{2-4} \cline{5-7}
\rule{0pt}{2.5ex} & Expt. & FSU & SDPF-MU & Expt. & FSU & SDPF-MU \\ \hline
\rule{0pt}{2.5ex}$11/2^- \rightarrow 9/2^-$  & $0.54^{+0.09*}_{-0.07}$   & 0.442 & 0.443 & $10^{+22*}_{-9}$ & 15.1 & 14.6 \\
\rule{0pt}{2.5ex}$13/2^- \rightarrow 11/2^-$ & $1.1^{+0.5 *}_{-0.3}$ & 0.258 & 0.411 & $30^{+18 *}_{-10}$ & 12.4 & 13.6 \\
\rule{0pt}{2.5ex}$15/2^- \rightarrow 13/2^-$ & $> 0.91^*$            & 0.345 & 0.409 & Unbounded$^*$      & 13.3 & 12.1 \\
\rule{0pt}{2.5ex}$11/2^- \rightarrow 7/2^-$  & \multicolumn{3}{c}{N/A}      & $< 44.8$         & 1.77 & 0.44 \\
\rule{0pt}{2.5ex}$13/2^- \rightarrow 9/2^-$  & \multicolumn{3}{c}{N/A}      & $< 39.6$         & 1.44 & 1.34 \\
\rule{0pt}{2.5ex}$15/2^- \rightarrow 11/2^-$ & \multicolumn{3}{c}{N/A}      & Unbounded        & 2.10 & 1.65 \\ \hline \hline
\rule{0pt}{2.5ex} \multirow{2}{*}{$J^{\pi}_i$} & \multicolumn{6}{c}{$B(M1; J \rightarrow J-1)$ $/$ $B(E2; J \rightarrow J-2)$}\\ \cline{2-7}
\rule{0pt}{2.5ex} & \multicolumn{3}{c}{W.u.} & \multicolumn{3}{c}{$(\mu_N/eb)^2$}\\ \hline
\rule{0pt}{2.5ex}$11/2^-$ & \multicolumn{3}{c}{$> 1.0 \times 10^{-2}$} & \multicolumn{3}{c}{$> 31$} \\
\rule{0pt}{2.5ex}$13/2^-$ & \multicolumn{3}{c}{$> 1.75 \times 10^{-2}$} & \multicolumn{3}{c}{$> 59$} \\
\end{tabular}
\end{ruledtabular}
\label{tab:b_table_29Al}
{\footnotesize \rule{0pt}{2.5ex}$^*$From Table \ref{tab:mixing_ratio_table}.}
\end{table*}

\par As with the $2^+_1$ level in $^{32}$Si, we found that some of our lifetime measurements in $^{29}$Al resulted in shorter values than previously reported. In this case, longer lifetimes were previously measured for several of the low-lying states in $^{26}$Mg($\alpha$,$p$) and $^{27}$Al($t$,$p$) data \cite{ekstrom_1974,beck_lifetime_1974}. Our lifetime measurements are more consistent with those reported in Ref.~\cite{sultana_2018}. Various sources of systematic error which could result in shortened lifetime values have been addressed, including target delamination, uncertainties in electronic stopping powers, and feeding corrections.

\section{Summary}

\par We have studied the high-spin structures of $^{32}$Si and $^{29}$Al using fusion-evaporation reactions and identified several new excited states, including candidates for the $4^+_1$, $6^+_1$ and $6^-_1$ levels of $^{32}$Si.  We have also performed lifetime measurements for all observed states. Our revised B($E2$; $2^+_1 \rightarrow 0^+_1$) value for $^{32}$Si suggests a gradual onset of deformation for the $N=18$ isotones approaching the `island of inversion' region.  Although the lower lying states are reproduced well by a variety of shell models, the predicted energies and transition strengths for higher spin and negative-parity states vary significantly depending on the interaction used, with particularly good agreement found using the recently developed FSU interaction.

\par In both nuclei, we find that the high spin ($J > 5$) yrast states have negative parity.  In $^{32}$Si, these negative-parity states mostly decay to a $5^-$ isomer, bypassing the $4^+_1$ state and leading to delayed feeding of the $2^+_1$ state.  In $^{29}$Al, the negative parity yrast states are members of a rotor-like band linked by strong \emph{M1} transitions, which mostly feeds the ground-state band.  The common thread is that both nuclei feature negative parity yrast states at high energy which decay to positive-parity states at low energy, leading to discrete normal and intruder structures in different excitation energy regimes.

\begin{acknowledgments}
The authors appreciate the support of the ISAC Operations Group at TRIUMF and the Simon Fraser University Electronics and Machine Shops. This work was supported by the Natural Sciences and Engineering Research Council of Canada, the Canadian Foundation for Innovation and the British Columbia Knowledge Development Fund. TRIUMF receives federal funding through a contribution agreement with the National Research Council of Canada. This work was also supported by the U.S. Department of Energy, Office of Science, Office of Nuclear Physics under Award No.~DE-SC0020451 (FRIB), and by the U.S. Department of Energy Grant No.~DE-FG02-93ER40789. FSU shell model calculations were performed using the computational facility of Florida State University, supported by Grant No.~DE-SC0009883 (FSU). P.C.S.~acknowledges financial support from SERB (India), CRG/2022/005167.
\end{acknowledgments}

%

\end{document}